\documentclass[a4paper]{nusthesis}

\usepackage{indentfirst} 

\usepackage{silence} 
\usepackage{bookmark}
\usepackage[
  backend=biber,
  style=ieee,
  citestyle=numeric,
  giveninits=true,
  sorting=nyt,
  maxbibnames=99,
  dashed=false,
  doi=false]{biblatex}
\DeclareFieldFormat*{title}{``#1''\newunitpunct} 
\usepackage{microtype} 
\usepackage{tabulary} 
\usepackage{hhline}

\usepackage{enumitem}

\usepackage{hyperref}

\usepackage{listings}

\usepackage{amsthm}
\theoremstyle{definition}
\newtheorem{defn}{Definition}

\theoremstyle{plain}
\newtheorem*{problem}{Problem}

\usepackage{mathtools}

\DeclarePairedDelimiter{\vbar}{\vert}{\vert}
\DeclarePairedDelimiter{\vbarbar}{\Vert}{\Vert}

\DeclareMathOperator*{\argmin}{arg\,min}
\DeclareMathOperator*{\argmax}{arg\,max}

\usepackage{interval}
\intervalconfig{soft open fences}

\usepackage[linesnumbered,ruled,vlined]{algorithm2e}
\SetAlgorithmName{Algorithm}{Algorithm}{Algorithm}
\newcommand{\AlgoFontSize}{\small} 
\IncMargin{0.5em}
\SetCommentSty{textnormal}
\SetNlSty{}{}{:}
\SetAlgoNlRelativeSize{0}
\SetKwInput{KwGlobal}{Global}
\SetKwInput{KwPrecondition}{Precondition}
\SetKwProg{Proc}{Procedure}{:}{}
\SetKwProg{Func}{Function}{:}{}
\SetKw{And}{and}
\SetKw{Or}{or}
\SetKw{To}{to}
\SetKw{DownTo}{downto}
\SetKw{Break}{break}
\SetKw{Continue}{continue}
\SetKw{SuchThat}{\textit{s.t.}}
\SetKw{WithRespectTo}{\textit{wrt}}
\SetKw{Iff}{\textit{iff.}}
\SetKw{MaxOf}{\textit{max of}}
\SetKw{MinOf}{\textit{min of}}
\SetKwBlock{Match}{match}{}{}

\usepackage{array}
\newcolumntype{L}[1]{>{\raggedright\let\newline\\\arraybackslash\hspace{0pt}}m{#1}}
\newcolumntype{C}[1]{>{\centering\let\newline\\\arraybackslash\hspace{0pt}}m{#1}}
\newcolumntype{R}[1]{>{\raggedleft\let\newline\\\arraybackslash\hspace{0pt}}m{#1}}

\usepackage{color}
\usepackage[usenames,dvipsnames]{xcolor}
\usepackage{soul}
\soulregister\cite7
\soulregister\ref7
\soulregister\pageref7
\soulregister\autoref7
\soulregister\eqref7

\renewcommand{\eqref}{Equation~\ref}

\newcommand*{\RootPicDir}{pic}
\newcommand*{\PicDir}{\RootPicDir}

\newcommand*{\SetPicSubDir}[1]{\renewcommand*{\PicDir}{\RootPicDir /#1}}
\newcommand*{\Pic}[2]{\PicDir /#2.#1}

\newcommand*{\RootExpDir}{exp}
\newcommand*{\ExpDir}{\RootExpDir}

\newcommand*{\SetExpSubDir}[1]{\renewcommand*{\ExpDir}{\RootExpDir /#1}}

\newcommand*{\BeforeCaptionVSpace}{1ex}

\usepackage{lipsum} 
\newcommand\attackname{ExploreADV}

\usepackage{derivative}

\usepackage{multirow}

\lstnewenvironment{code}[1][]{\lstset{#1}}{}

\providecommand{\keywords}[1]
{
  \small	
  \textbf{\textit{Keywords---}} #1
}
\usepackage[normalem]{ulem}
\begin{document}

\title{\attackname{}: Towards exploratory attack for Neural Networks}

\author{LUO Tianzuo}
\degree{Master of Computing}
\field{Computer Science}
\degreeyear{2022}
\supervisor{Associate Professor KHOO Siau Cheng}


\maketitle


\newcommand{\ksc}[2]{\sout{#1}~\textcolor{blue}{[#2]}}
\newcommand{\khoo}[1]{\textcolor{blue}{[KHOO: }\textcolor{blue}{#1}\textcolor{blue}{]}}

\newcommand{\zyy}[1]{\textcolor{red}{[YUYI: }\textcolor{red}{#1}\textcolor{red}{]}}

\newcommand{\reply}[1]{\textcolor{cyan}{[Reply: }\textcolor{cyan}{#1}\textcolor{cyan}{]}}

\newcommand{\highlight}[1]{\textcolor{cyan}{#1}}

\begin{frontmatter}
  \tableofcontents 
  \begin{abstract}

Although deep learning has made remarkable progress in processing various types of data such as images, text and speech, they are known to be susceptible to adversarial perturbations: perturbations specifically designed and added to the input to make the target model produce erroneous output. Most of the existing studies on generating adversarial perturbations attempt to perturb the entire input indiscriminately. In this paper, we propose \attackname{}, a general and flexible adversarial attack system that is capable of modeling regional and imperceptible attacks, allowing users to explore various kinds of adversarial examples as needed. We adapt and combine two existing boundary attack methods, DeepFool and Brendel\&Bethge Attack, and propose a mask-constrained adversarial attack system, which generates minimal adversarial perturbations under the pixel-level constraints, namely ``mask-constraints''. We study different ways of generating such mask-constraints considering the variance and importance of the input features, and show that our adversarial attack system offers users good flexibility to focus on sub-regions of inputs, explore imperceptible perturbations and understand the vulnerability of pixels/regions to adversarial attacks. We demonstrate our system to be effective based on extensive experiments and user study. 


\end{abstract}

  \listoffigures
  \listoftables
\end{frontmatter}

\SetPicSubDir{ch-Intro}

\chapter{Introduction}
\vspace{2em}

In recent years, deep learning has become a critical role in a variety of domains such as computer vision \cite{voulodimos2018deep}, natural language processing \cite{otter2020survey}, speech and audio processing \cite{purwins2019deep}, etc. It has made significant breakthroughs, especially in the tasks of image classification \cite{krizhevsky2012imagenet,he2016deep,huang2017densely}, segmentation \cite{long2015fully,he2017mask} and object detection \cite{girshick2014rich,girshick2015fast,ren2015faster}, where deep learning has achieved high accuracy and even exceeded human performance. Szegedy et al. \cite{szegedy2013intriguing} found an intriguing property of deep neural networks, that it is possible to arbitrarily
change the network’s prediction by applying an imperceptible and non-random perturbation to the test image. In this work, we propose a novel system to study such examples, also known as ``adversarial examples''.

\section{Deep learning}

Deep learning is part of a broader family of machine learning methods \cite{lecun2015deep}. It uses artificial neural network composed of a large number of neurons with activation functions to perform representation learning of data. Deep neural network can automatically learn the explicit and implicit features of the original data without relying on expert knowledge.

\begin{figure}[h]
  \centering
  \includegraphics[width=.8\linewidth]{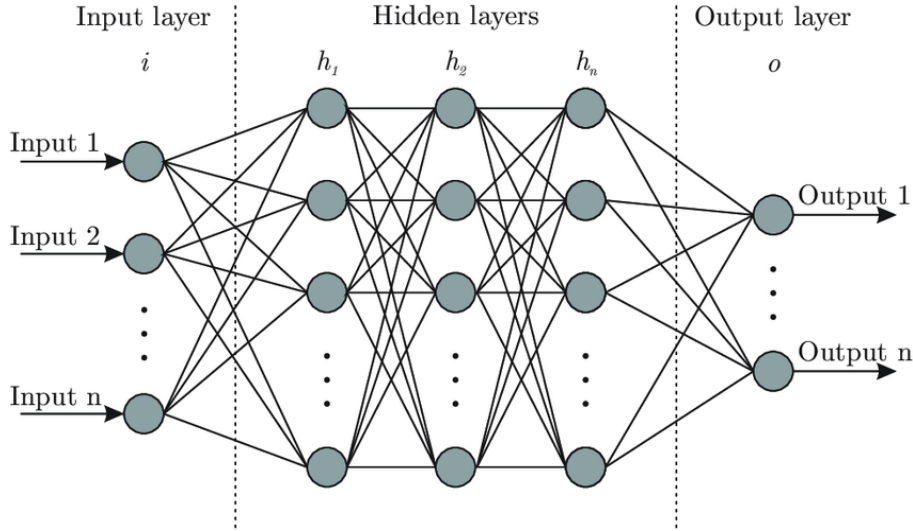}
  \vspace{\BeforeCaptionVSpace}
  \caption{Artificial neural network architecture \cite{bre2018prediction}}
  \label{intro:fig:ANN}
\end{figure}

A typical artificial neural network architecture is shown in \autoref{intro:fig:ANN}. Each of the neurons receives input signal from previous layer and performs weighted connection, then processes the output of the neuron through an activation function and transmits the signal to next layer, thus constructing a deep neural network structure. It can be formally expressed as shown in \autoref{eq:intro:1}.
\begin{align}
  y = h_n(...h_2(w_2 \cdot h_1(w_1 \cdot x + b_1) + b_2)) \label{eq:intro:1}
\end{align}
where $x$ and $y$ are the input and output of the network, $w_i$, $b_i$ and $h_i$ are the respective weights, biases and activation functions in the $i^{th}$ layer of the network, i = 1, 2, ..., n, where n is the number of hidden layers in the network.

Even though deep neural network has achieved remarkable results by simulating the structure of human brain neural network, the way deep neural networks work is still quite different from human cognition and lack of interpretability, making it difficult to guarantee the credibility of its output. 

The growing use of deep neural networks has raised concerns about their security and reliability. Szegedy et al. \cite{szegedy2013intriguing} found that deep neural networks are highly vulnerable to image samples with specific perturbations, and called such image samples with adversarial perturbations as ``adversarial examples''. 

\section{Adversarial Example}

An adversarial example refers to the input sample formed by adding specifically designed perturbations to an original sample, which can make the well-trained deep learning model give erroneous outputs. Specifically, In the field of computer vision, an adversarial example is usually an image formed by adding slight perturbations to the input image that are difficult to be perceived by human vision, resulting in incorrect prediction from the model, e.g. identifying a panda as a gibbon. 

\begin{figure}[!t]
  \centering
  \includegraphics[width=.8\linewidth]{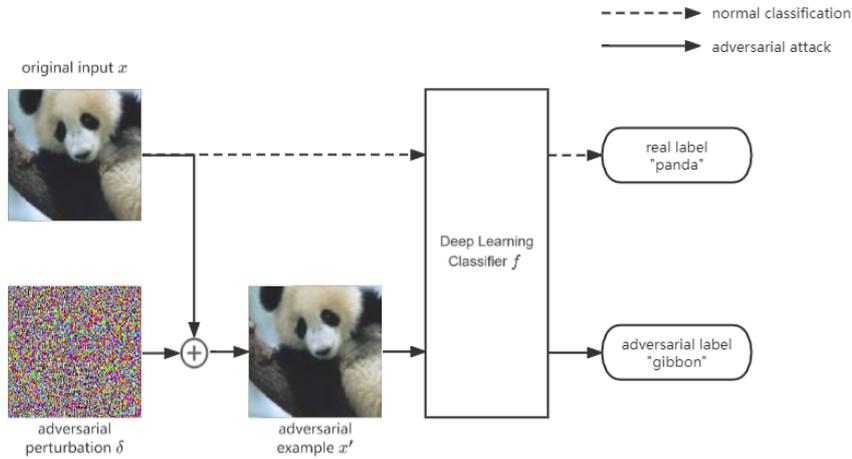}
  \vspace{\BeforeCaptionVSpace}
  \caption{Process of Adversarial Attack}
  \label{intro:fig:flow} 
\end{figure}

Adversarial attack is the procedure of generating adversarial examples in order to fool a deep learning model. \autoref{intro:fig:flow} demonstrates the process of adversarial attacks. A variety of attack algorithms have been proposed to generate adversarial examples \cite{szegedy2013intriguing, goodfellow2014explaining, madry2017towards, moosavi2016deepfool,carlini2017towards}. Without loss of generality, we formally define an adversarial example in the context of image classification problem.

\begin{defn}[\textbf{Adversarial Example}]
\label{def:adv}
Given an input image $x \in \mathbb{R}^{d}$, and a score-based image classifier $f: \mathbb{R}^{d} \mapsto \mathbb{R}^{K}$ that maps $x$ to a set of K labels $S = \{1, 2, ..., K\}$ according to:
\begin{align}
  \begin{split}
  \hat{y}(x) = \argmax_{k \in S}f_k(x)
  \end{split} \label{eq:intro:2}
\end{align}
where $f_k(x)$ is the score function for label $k \in S$, $\hat{y}(x) \in S$ is the predicted label for input $x$. 

The collection of adversarial examples with respect to $x$ and $f$ is defined as:
\begin{align}
  \begin{split}
  \{ x' ~|~ d(x, x') < \epsilon,~ \hat{y}(x') \neq \hat{y}(x) \}
  \end{split} \label{eq:intro:3}
\end{align}
where $d(x, x')$ is the distance (a measure to be discussed in \autoref{ssec:lp}) between the adversarial example and the original input, and will be bounded by a small predefined constant $\epsilon$. Each adversarial example $x'$ can be considered as a combination of the original image $x$ and an adversarial perturbation $\delta$, i.e., $x' = x+\delta$. When using $L_p$-norms as distance metric, $d(x, x')=\vbarbar{x'-x}_p=\vbarbar{\delta}_p<\epsilon$.
\end{defn}

In order to keep the adversarial image perceptually close to the original image, good distance metrics to measure the perceptual similarity between two images are important. Ideally, smaller distance represents closer similarity with respect to human perception. As it is difficult to quantitatively measure human perception, in many classical adversarial attack algorithms \cite{szegedy2013intriguing,goodfellow2014explaining,papernot2016limitations,moosavi2016deepfool,carlini2017towards}, $L_p$-norm distance is applied. 

\subsection{$L_p$-norm distance}
\label{ssec:lp}
To measure the distance between an image $x$ and its adversarial image $x'$, $L_p$-norm distance is defined by the $L_p$-norm of the pixel value difference $\vbarbar{x' - x}_{p}$, i.e., the $L_p$-norm of the adversarial perturbation: $\vbarbar{\delta}_{p}$. The definition of $L_p$-norm is shown in \autoref{def:lp}.

\begin{defn}[\textbf{$L_p$-norm}]
\label{def:lp}
Given a vector $\delta = (\delta_1, \delta_2, ..., \delta_n)$ in the n-dimensional real vector space $\mathbb{R}^{n}$, and a real number $p \geq 1$, the $L_p$-norm of $\delta$ is defined by:
\begin{align}
  \vbarbar{\delta}_{p} = (\sum_{i=1}^{n}\vbar{\delta_{i}}^{p})^{\frac{1}{p}} \label{eq:intro:5}
\end{align}
\end{defn}

In practise, $L_0$, $L_1$, $L_2$, and $L_\infty$-norm distances are commonly used:

\begin{itemize}
\item $L_1$ (Manhattan distance): $\vbarbar{\delta}_1 = \sum_{i=1}^{n}\vbar{\delta_{i}}$
\item $L_2$ (Euclidean distance): $\vbarbar{\delta}_2 = \sqrt{\sum_{i=1}^{n}\vbar{\delta_{i}}^2}$
\item $L_\infty$ (Chebyshev distance): $\vbarbar{\delta}_\infty = \max_{i}\vbar{\delta_{i}}$
\end{itemize}

$L_0$-norm is special, it's defined as $\vbarbar{\delta}_0 = \sum_{i=1}^{n}\{1~|~\delta_i \neq 0\}$, which counts the number of non-zero pixel-value differences. It is actually not a norm because it does not satisfy absolute homogeneity\footnote{Given a vector space $V$, a function $f: X \mapsto \mathbb{R}$ satisfies absolute homogeneity if $f(\lambda v)=\vbar{\lambda}f(v)$ for all $v \in V$ and $\lambda \in \mathbb{R}$, where $\vbar{\lambda}$ denotes the absolute value of the scalar $\lambda$ \cite{pugh2002real}.}.

From the definition, it can be noticed that $L_p$-norm distance is only related to the pixel value differences $\delta$, it is not affected by the actual pixel values in the clean image $x$ or its adversarial image $x'$.

\subsection{Minimal Adversarial Perturbation}

More recent attention has focused on finding the \textit{minimal adversarial perturbation}, also known as the robustness of model at point $x$ \cite{moosavi2016deepfool}. They typically use the $L_p$-norm distances as the distance metric, and try to find the minimal perturbation necessary to change the prediction of the model. The minimal adversarial perturbation with respect to the $L_p$-norm is defined as:
\begin{align}
  \begin{split}
  \argmin_{\delta}\vbarbar{\delta}_{p}, ~ \delta \in \{ \delta ~|~  \hat{y}(x+\delta) \neq \hat{y}(x) \}
  \end{split} \label{eq:intro:4}
\end{align}

The optimization problem in \autoref{eq:intro:4} is NP-complete for non-linear and non-convex classifiers \cite{katz2017reluplex}. In practice, it is often approximated by different attack algorithms, either by using some heuristics \cite{goodfellow2014explaining,moosavi2016deepfool, croce2020minimally} or by solving minimization problems \cite{szegedy2013intriguing,carlini2017towards}.

Specifically, when considering minimal adversarial perturbation with respect to $L_\infty$-norm, we call the perturbation size the \textit{robust radius} \cite{zhai2020macer}, as defined in \autoref{def:robust}.

\begin{defn}[\textbf{Robust Radius}]
\label{def:robust}
Given an input image $x \in \mathbb{R}^{d}$, and a score-based image classifier with prediction function $\hat{y}(x)$. The robust radius $r_* \in \mathbb{R}$ of the classifier on $x$ is defined as:
\begin{align}
  \begin{split}
  &r_* = min(r) \\
  &\text{ s.th. } \exists x'\: \hat{y}(x') \neq \hat{y}(x) \: \text{and} \: \vbar{x'_i-x_i} \leq r\: \text{for i = 1, ..., d}
  \end{split} \label{eq:intro:robust}
\end{align}
\end{defn}

\section{Limitation of previous work}
\label{sec:intro:limitation}

\begin{figure}[h]
  \centering
  \includegraphics[width=.8\linewidth]{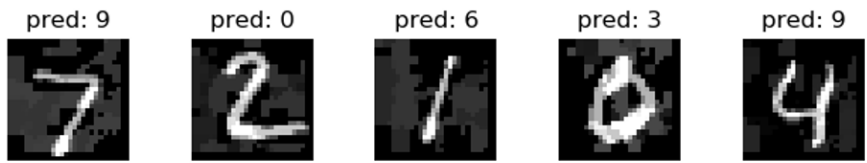}
  \vspace{\BeforeCaptionVSpace}
  \caption[Adversarial Images on MNIST dataset]{\textbf{Adversarial Images on MNIST dataset.} These adversarial images are supposed to represent the digits 7, 2, 1, 0 and 4, and are predicted as 9, 0, 6, 3 and 9, yet they look blurred and unrealistic.}
  \label{intro:fig:blurred}
\end{figure}

\textbf{Existing attacks are not perceptually constrained.} While most adversarial attack algorithms try to find adversarial perturbation with small $L_p$-norms, it is argued that using $L_p$-norms to measure the perceptual similarity between two images is neither necessary nor sufficient \cite{sharif2018suitability}. When using $L_p$-norms as the distance metric, it implies the assumption that perturbations on different pixels in an image are equally important for human eyes. However, as Liu et al. \cite{liu2010just} suggests, perturbations become less perceptible in the regions with high spatial variation, and more perceptible in smooth regions. As shown in \autoref{intro:fig:blurred}, the adversarial images found by some existing adversarial attack algorithms \cite{moosavi2016deepfool}, while aiming to have small $L_p$-norm, appear perceptually blurred and unrealistic with perceptible noise.

Studies also showed that the adversarial examples found by some existing methods neither faithfully simulate physical objects nor resemble natural images \cite{lu2017no, yosinski2015understanding}. 

To develop methods to find adversarial examples perceptually closer to the original image, better perceptual distance metrics are needed to evaluate the effective of the methods. There are other image similarity distance metrics proposed, such as CIEDE2000 \cite{luo2001development} and SSIM \cite{wang2004image}, that can supplement the $L_p$-norms for measuring perceptual similarity. Details of the metrics can be found in \autoref{append:metric}.

\textbf{Existing attacks are not suitable for modeling real-world threats.} Recent research also suggests that adversarial examples can be generalized to the real world. Sharif et al. \cite{sharif2016accessorize} showed that face recognition systems can be fooled by people wearing adversarially constructed eyeglass frames. Brown et al. \cite{brown2017adversarial} create a method to generate ``adversarial patches'' that can be printed and added to the scene to fool a classifier. While such ``physical-world'' attacks may seem practical for real-world ML systems, it is currently not suitable to be modeled by most of the existing attacks which aim to modify the whole image indiscriminately — ``physical-world'' attacks on an entire scene is usually not feasible.
As a result, some attack methods seek to perturb only few pixels \cite{papernot2016limitations} or a small region in the image \cite{sharif2016accessorize}. Nevertheless, these attacks often do not restrict themselves to imperceptible perturbations, the resulting adversarial perturbations are often clearly visible.

\vspace{2em}
To overcome these limitations and take advantage of the power of existing adversarial attack methods, a natural approach is to properly constrain the adversarial perturbations during the attack. For example, the perturbation can be constrained to be:
\begin{itemize}
\item \textbf{Regional.} Keep pixels unperturbed outside the target region.
\item \textbf{Imperceptible.} Reduce perturbations of pixels in regions where such perturbations are more perceptible.
\end{itemize}

\section{Our work}
\begin{figure}[h]
  \centering
  \includegraphics[width=\linewidth]{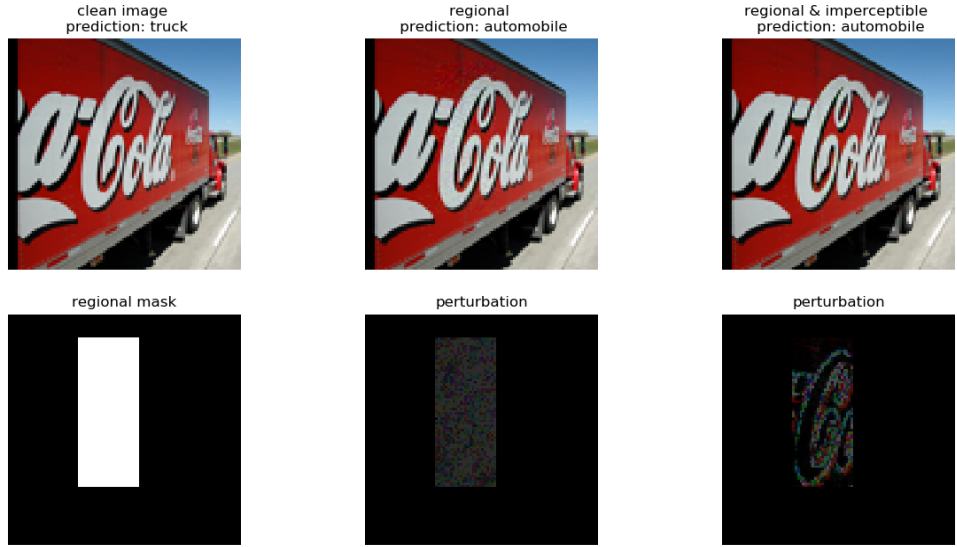}
  \vspace{\BeforeCaptionVSpace}
  \caption[Regional and Imperceptible sticker on a truck.]{\textbf{Regional and Imperceptible sticker on a truck.} We illustrate the adversarial examples found by our system by adding perturbation to a small region of a truck. From left to right. \textit{left column} - original image, and the regional mask indicating the region to apply attack (white area) where the black area remains unperturbed. \textit{middle column} - normal adversarial examples that is perturbed regionally, and the adversarial perturbations, \textit{
 right column} - imperceptible adversarial examples that is perturbed regionally, and the adversarial perturbations.} 
  \label{intro:fig:example}
\end{figure}

In this paper, we propose \attackname{}, a general and flexible adversarial attack system that is capable of modeling regional and imperceptible attacks. A novel type of constraint, namely ``mask-constraint'' is proposed in our system. The system allows users to explore different types of threat models based on their interest, e.g. they can decide for an image the region they want to focus on, whether they want the perturbation to be imperceptible, and how many pixels / how large a region they want to perturb. For example, to test the reliability of the vision model on a self-driving system, one may want to know if a maliciously designed sticker on a truck would deceive the system, and whether such sticker can be designed imperceptible so that people would not notice any anomaly before a potential accident happens. Such exploration are possible in our system, as shown in \autoref{intro:fig:example}.

We propose an idea of mask-constraint in our system to enable such flexibility, as defined in \autoref{def:mask}.

\begin{defn}[\textbf{Mask-constraint}]
\label{def:mask}
Given a clean image $x=(x_1, x_2, ..., x_d)$ with $d$ pixels, a mask-constraint contains a set of non-negative constant constraints $\mathcal{E}=(\epsilon_1, \epsilon_2, ..., \epsilon_d)$, where $\epsilon_i \in [0, 1]$ is the constraint on the $i^{th}$ pixel $x_i$. Each $\epsilon_i$ indicates the maximum allowable absolute perturbation on $x_i$, where 0 means no perturbation allowed. An adversarial image $x'=(x'_1, x'_2, ..., x'_d)$ found under the mask-constraint is limited in the closed interval $x_i-\epsilon_i \leq x'_i \leq x_i+\epsilon_i$ for each pixel $x'_i$.
\end{defn}

With the mask-constraints in our system, we formulate  the problem of finding adversarial perturbation as:

\begin{problem}[\textbf{Adversarial perturbation under mask-constraint}]
\label{prob:adv_mask}
Given an image classification neural network $N$ with prediction function $\hat{y}(x)$, an image $x = (x_1, x_2, ..., x_d)$, a mask-constraint $\mathcal{E} = (\epsilon_1, \epsilon_2, ..., \epsilon_d)$ on each pixel of $x$, and a real number $p \geq 1$. Find a perturbation $\delta = (\delta_1, \delta_2, ..., \delta_d)$ satisfying $\{\delta_i~|~-\epsilon_i \leq \delta_i \leq \epsilon_i\}$, such that when adding the perturbation to the image, the model's prediction on the resulting image $x' = x+\delta$ is different from its prediction on the original image: $\hat{y}(x') \neq \hat{y}(x)$.  Moreover, the $L_p$-norm of the perturbation $\vbarbar{\delta}_p$ is minimized.
\end{problem}

In this thesis, we propose a novel approach to solve this problem. We adapt and combine two existing adversarial attack methods, DeepFool \cite{moosavi2016deepfool} and Brendel\&Bethge \cite{brendel2019accurate} Attack, which will be introduced in detail in \autoref{ch:review}, and adapt them to work under our mask-constraint setting. DeepFool is first applied to yield a preliminary adversarial example under the mask-constraint. If a preliminary adversarial example is found, it is used as a starting point for Brendel\&Bethge attack, which can then minimize the $L_p$-norm of the perturbation under the same mask-constraint, if DeepFool fails to find an adversarial example, the system terminates and returns no result, an adversarial example might not exist or might be found with more iterations of DeepFool. We later show that the integration of the mask-constraint makes regional and imperceptible adversarial perturbations possible in our system.

To facilitate users to automatically generate imperceptible perturbations and select pixels/regions to perturb, we study two types of maps reflecting the variance and importance of pixels in this work:
\begin{description}
  \item[Variance map] The variance map measures the spatial variation of the image by calculating the variance of pixel values in a small neighbourhood. It is helpful when imperceptibility is desired, as perturbations in regions with high spatial variation are less perceptible than those in smooth regions. Our system use a variance map based method to generate constraints on pixels according to their variance, allowing less perturbation for pixels in regions with low variance.
  \item[Importance map] The importance map measures the importance of each pixel to changing the prediction of the classifier. It is helpful when there is a need to perturb only a subset of all pixels. Our system use an importance map based method to estimate the vulnerability of pixels/regions in the image to adversarial attacks, and select pixels/regions with high vulnerability to apply perturbations.
\end{description}
It is worth noticing that there are many ways to generate these maps, and our methods are not restricted to any single implementation of them. There has been a few attempts to make use of such variance map \cite{croce2019sparse} and importance map \cite{bai2021inconspicuous}, but we adapt or improve their methods in this work.

In this work, we only consider $L_\infty$-norm as it is commonly used to access model robustness \cite{singh2019abstract}, so our system can be used to estimate the robust radius under the mask-constraint. Yet, our method is orthogonal to the selection of $L_p$-norms and can be easily extended to other norms.

Our main contributions can be summarized as follows:

\begin{itemize}

  \item We propose a novel adversarial attack system with mask-constraints, which is more general than existing attacks because it can limit the adversarial perturbation to any sub-region of the whole image, and limit the perturbation magnitude on any pixel of the image independently.

  \item We adopt and combine two existing adversarial attack methods, DeepFool and Brendel\&Bethge Attack, and show that the resulting method generates adversarial perturbations with small $L_\infty$ norm that are comparable to the adversarial perturbations generated by the state of the art $L_\infty$ attack methods \cite{croce2020minimally, pintor2021fast}. 

  \item We study different ways to automatically generate mask-constraints in our system by considering the variance and importance of the pixels in the image, which provides the user with much flexibility to explore various kinds of adversarial examples conveniently, to generate regional and imperceptible adversarial perturbations as they need.
  
  \item We suggest ways to enhance variance map and importance map based method. We propose to adaptively loosen the variance map based mask-constraint to generate imperceptible perturbations for models with different robustness, and to add a correction coefficient to the importance map to better estimate pixel vulnerability.
\end{itemize}

\section{Thesis Synopsis}

The rest of this thesis is organized as follows. 
In \autoref{ch:review}, we conduct a literature review on related works. \autoref{ch:method} provides details of our method. Extensive experimental results are depicted in \autoref{ch:evaluation}. We conclude the entire thesis as well as discuss further directions for future research in \autoref{ch:concl}.

\SetPicSubDir{ch-Review}

\chapter{Related Work}
\label{ch:review}
\vspace{2em}

In this section, we review some prior work of adversarial attacks that are related to ours.

\section{DeepFool attack}
Considering that deep neural network is extremely vulnerable to adversarial examples, Moosavi-Dezfooli et al. \cite{moosavi2016deepfool} proposed a method called DeepFool, which aims to calculate the minimal perturbation with respect to $L_2$-norm (extendable to other $L_p$-norms) necessary to change the classifier’s decision, as shown in \autoref{eq:review:df:1}. 
\begin{align}
  \begin{split}
  \delta_* &= \argmin_{\delta} \vbarbar{\delta}_2 \text{ subject to } \hat{y}(x + \delta) \neq \hat{y}(x)
  \end{split} \label{eq:review:df:1}
\end{align}
where $x$ is an image and $\delta_*$ is the minimum perturbation. $\hat{y}$ is the prediction function as in \autoref{eq:intro:2}. 

Since a multi-class classifier can be viewed as an aggregation of binary classifiers, they first introduced an efficient algorithm to find adversarial examples for binary classifiers, then extended it to the multi-class case.

\subsection{Binary classifier}

For the binary classifier, assume $\hat{y}(x) = sign(f(x))$, where $f: \mathbb{R}^{d} \mapsto \mathbb{R}$ represents an arbitrary scalar-valued image classifier, $sign(\cdot)$ is the sign function that extracts the sign of a real number.

\subsubsection{Affine classifier}

Given a binary affine classifier $f(x) = \omega^Tx+b$, the corresponding affine hyperplane $\mathcal{F} = \{x | \omega^Tx+b = 0\}$. Then the minimum perturbation $\delta_*$ to change the classifier’s prediction on the original sample $x^0$ is equal to the orthogonal projection of $x^0$ onto the affine hyperplane $\mathcal{F}$, which is given by the closed-form formula: 
\begin{align}
  \begin{split}
  \delta_*(x^0) =& \argmin_{\delta} \vbarbar{\delta}_2 \\
  &\text{ subject to } sign(f(x^0 + \delta)) \neq sign(f(x^0)) \\
  =& \frac{f(x^0)}{\vbarbar{\omega}_2^2}\omega
  \end{split} \label{eq:review:df:2}
\end{align}

\subsubsection{General classifier}

When $f$ is a general binary differentiable classifier, DeepFool adopts an iterative procedure to estimate the minimum perturbation $\delta$. At each iteration $i$, $f$ is linearized around the current point $x^i$, and the minimal perturbation $\delta_i$ is computed as:
\begin{align}
  \begin{split}
  &\argmin_{\delta^i} \vbarbar{\delta^i}_2 \\
  &\text{such that } f(x^i) + \nabla f(x^i)^T \delta^i = 0
  \end{split} \label{eq:review:df:3}
\end{align}

\subsection{Multi-class classifier}

For the multi-class classifier, assume $\hat{y}(x) = \argmax_{k \in S}f_k(x)$, where $f: \mathbb{R}^{d} \mapsto \mathbb{R}^{K}$ represents an arbitrary score-based image classifier, $f_k(x)$ is the score function of $f(x)$ for corresponds to label $k$.

\subsubsection{Affine classifier}
Given an affine classifier $f(x) = \omega^Tx+b$. The minimum perturbation that spoofs the classifier can be overridden in the following way:
\begin{align}
  \begin{split}
  \delta_*(x^0) &= \argmin_{\delta} \vbarbar{\delta}_2 \\
  &\text{ s.th. } \exists k: f_{\hat{y}(x^0)}(x^0 + \delta) \leq f_k(x^0 + \delta)\\
  &=\frac{\vbar{f_l(x^0) - f_{\hat{y}(x^0)}(x^0)}} {\vbarbar{\omega_l - \omega_{\hat{y}(x^0)}}_2^2}(\omega_l - \omega_{\hat{y}(x^0)})
  \end{split} \label{eq:review:df:4}
\end{align}
where $l$ is the class different from $\hat{y}(x^0)$ with the closest hyperplane of the boundary, as shown in \autoref{eq:review:df:5}.
\begin{align}
  \begin{split}
  l = \argmin_{k \neq \hat{y}(x^0)} \frac{\vbar{f_k(x^0) - f_{\hat{y}(x^0)}(x^0)}} {\vbarbar{\omega_k - \omega_{\hat{y}(x^0)}}_2}
  \end{split} \label{eq:review:df:5}
\end{align}

\subsubsection{General classifier}

In general case of multi-class classifiers, DeepFool attack pushes the original sample toward the decision boundary with each round of perturbation until it crosses the decision boundary to form an adversarial example or reach the maximum allowable iterations, as shown in \autoref{Review:algo:DeepFool}. 

\begin{algorithm}[h]
\AlgoFontSize
\DontPrintSemicolon

\KwIn{image $x$, classifier $f$, maximum iterations $max\_iter$}
\KwOut{perturbation $\hat{\delta}$}
Initialize $x^0 \gets x, i \gets 0$\;
\While{$\hat{y}(x^i) = \hat{y}(x^0)$ and $i < max\_iter$}{
\For{$k \neq \hat{y}(x^0)$}{
  $f'_k \gets f_k(x^i) - f_{\hat{y}(x^0)}(x^i)$\;
  $\omega'_k \gets \nabla f'_k$\;
}
$\hat{l} \gets \argmin_{k \neq \hat{y}(x^0)}{\frac{\vbar{f'_k}}{\vbarbar{\omega'_k}_2}}$\; \label{algo:l}
$\delta^i \gets \frac{\vbar{f'_{\hat{l}}}}{\vbarbar{\omega'_{\hat{l}}}_2^2}\omega'_{\hat{l}}$\; \label{algo:delta}
$x^{i+1} \gets x^i + \delta^i$\;
$i \gets i + 1$\;
}
\Return $\hat{\delta} = \sum_{i}\delta^i$\;

\caption{DeepFool: multi-class case}
\label{Review:algo:DeepFool}
\end{algorithm}

\subsection{Extend to $L_p$-norm}

Although DeepFool was proposed based on $L_2$-norm, it can simply be adapted to find minimal adversarial perturbations for any $L_p$-norms ($p \in \{ \infty \} \cup [1, \infty)$), by substituting the update steps in \autoref{algo:l} and \autoref{algo:delta} in \autoref{Review:algo:DeepFool} with the following steps respectively:
\begin{align}
&\hat{l} \gets \argmin_{k \neq \hat{y}(x^0)}{\frac{\vbar{f'_k}}{\vbarbar{\omega'_k}_q}} \label{eq:review:df:6}\\
&\delta^i \gets \frac{\vbar{f'_{\hat{l}}}}{\vbarbar{\omega'_{\hat{l}}}_q^q}\vbar{\omega'_{\hat{l}}}^{q-1}\odot sign(\omega'_{\hat{l}}) \label{eq:review:df:7}
\end{align}
where $\odot$ is the element-wise multiplication, $q=\frac{p}{p-1}$. 

Note that for $p=2$ as in \autoref{Review:algo:DeepFool}, the steps in \autoref{algo:l} and \autoref{algo:delta} are also equivalent to the steps in \autoref{eq:review:df:6} and \autoref{eq:review:df:7}. When $p=2$, $q=\frac{2}{2-1}=2$, \autoref{eq:review:df:6} matches \autoref{algo:l} of \autoref{Review:algo:DeepFool}, and the right half of \autoref{eq:review:df:7} can be written as $\vbar{\omega'_{\hat{l}}} \odot sign(\omega'_{\hat{l}})$, which equals to $\omega'_{\hat{l}}$ in \autoref{algo:delta} of \autoref{Review:algo:DeepFool}. 

For $L_\infty$-norm, the steps become:
\begin{align}
&\hat{l} \gets \argmin_{k \neq \hat{y}(x^0)}{\frac{\vbar{f'_k}}{\vbarbar{\omega'_k}_1}} \\
&\delta^i \gets \frac{\vbar{f'_{\hat{l}}}}{\vbarbar{\omega'_{\hat{l}}}_1}sign(\omega'_{\hat{l}})
\end{align}

\subsection{Summary}

DeepFool uses a local linear approximation of the classifier to estimate the optimal step towards the decision boundary. Compared with FGSM \cite{goodfellow2014explaining} attack, DeepFool attack generates adversarial examples with smaller perturbation, and close to the decision boundary.

\section{Brendel \& Bethge (BB) attack}

\begin{figure}[!t]
  \centering
  \includegraphics[width=.8\linewidth]{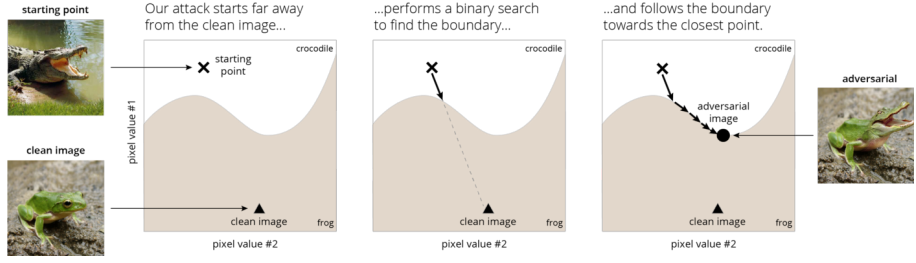}
  \vspace{\BeforeCaptionVSpace}
  \caption{Schematic of Brendel \& Bethge (BB) attack \cite{brendel_2020}}
  \label{review:fig:BB}
\end{figure}

Brendel et al. \cite{brendel2019accurate} developed Brendel \& Bethge (BB) attack, a new set of gradient-based adversarial attacks to find local minimal adversarial examples. The scheme of the attack is shown in \autoref{review:fig:BB}. It starts from a randomly-drawn adversarial example that is far away from the clean image (left in \autoref{review:fig:BB}), and first performs a 10-step binary search to reach the decision boundary (middle in \autoref{review:fig:BB}), then walk along the decision boundary to minimize the $L_p$ distance to the clean image (right in \autoref{review:fig:BB}). At each step $k$, they solve a constrained optimization problem to find the optimal descent step $\delta^k$ within a trust radius $r$, and add it to the current image $\tilde{x}^{k-1}$. Finally, $\tilde{x}^{k}$ is returned, with the $L_p$-norm of the adversarial perturbation minimized.

The constrained optimization problem is defined as:
\begin{align}
  \begin{split}
  &\argmin_{\delta} \vbarbar{x-\tilde{x}^{k-1}-\delta^k}_p \\
  &\text{ s.th. } 0 \leq \tilde{x}^{k-1}+\delta^k \leq 1 \land {b^k}^T\delta^k=c^k \land \vbarbar{\delta^k}^2_2 \leq r
  \end{split} \label{eq:BB}
\end{align}
where $x$ is the clean image, $\tilde{x}^{k-1}$ is the perturbed image after step k-1, $\delta^k$ is the perturbation for step k, $b^k$ is the direction of the local decision boundary, $c^k$ is the distance to the decision boundary. The decision boundary is defined by a differentiable equality constraint $adv(\tilde{x}) = 0$, where $adv(\cdot)$ is the adversarial criterion function. 

Let $f_t(\tilde{x})$ be the log-probability for label $t$ on the current input $\tilde{x}$, $y$ being the true label for the clean image $x$, then for targeted attack:
\begin{align}
  \begin{split}
  adv(\tilde{x}) = f_y(\tilde{x}) - f_t(\tilde{x})
  \end{split}
\end{align}
When $adv(\tilde{x}) = 0$, the log-probability of $y$ is equal to the log-probability $t$. For untargeted attack:
\begin{align}
  \begin{split}
  adv(\tilde{x})= \min_{t, t \neq y}(f_y(\tilde{x}) - f_t(\tilde{x}))
  \end{split}
\end{align}
When $adv(\tilde{x}) = 0$, the log-probability of $y$ is equal to the log-probability of the class with the highest log-probability among the other classes.

\section{Imperceptible attack}
\label{sec:review:imperc}

Luo et al. \cite{luo2018towards} discovered, by investigating the human visual system, that human eyes are more sensitive to perturbation in flat areas than textured areas. Therefore, the texture features around a pixel should be taken into consideration when attempting to add adversarial perturbation to the pixel. Texture features of an image $x$ can be quantified by the notion of {\em variance}, as shown in \autoref{eq:review:imper:1}.
\begin{align}
  SD(x_i) = \sqrt{\frac{\sum_{x_k \in S_i}(x_k-\mu)^2}{n^2}}
  \label{eq:review:imper:1}
\end{align}
where $SD(x_i)$ represents the standard deviation of the pixel values among an n × n neighborhood $S_i$ of pixel $x_i$. Considering the impact of perturbation intensity on human vision, they introduced the notion of ``perturbation sensitivity'' to measure how much “attention” will be drawn by adding per ``unit'' perturbation on a pixel. 
\begin{align}
  Sen(x_i) = 1/SD(x_i)
  \label{eq:review:imper:2}
\end{align}

They further defined a distance metric based on this notion of ``perturbation sensitivity'', as shown in \autoref{eq:review:imper:3}.
\begin{align}
  d(x, x') = \sum_{i = 1}^{m} \vbar{\delta_i} * Sen(x_i)
  \label{eq:review:imper:3}
\end{align}
where $d(x, x')$ denotes the distance between the adversarial example $x'=(x'_1, x'_2, ..., x'_m)$ and the original one $x=(x_1, x_2, ..., x_m)$, $m$ is the total number of pixels and $\vbar{\delta_i}$ is the intensity of perturbation on pixel $x_i$. 

They then proposed a targeted attack, which attempts to mis-classify an image into a specific target class. They used a greedy algorithm to select pixels with lower sensitivity and higher impact (larger gradient) in order to maximize the gap between the probability of the target class and the maximal probability of all other classes, and perturb the pixels in an iterative manner under certain constraint $d(x, x') \leq d_{max}$.

Croce et al. \cite{croce2019sparse} proposed sparse and imperceivable adversarial attacks, with locally adaptive component-wise constraints on the perturbation. They defined the constraint on each pixel as:
\begin{align}
  \sigma_{ij} = \kappa\sqrt{min(\sigma_{ij}^{(x)}, \sigma_{ij}^{(y)})}
  \label{eq:review:imper:4}
\end{align}
where $\kappa$ is a hyper-parameter set by users, $\sigma_{ij}^{(x)}$, $\sigma_{ij}^{(y)}$ are the standard deviation of each color channel in x- and y-axis directions with the two immediate neighboring pixels and the original pixel, $i \in [1, d]$ represents each of the $d$ pixels, $j \in [1, 3]$ represents each color channel. Given the constraints on each pixel, the adversarial example generated under the constraint can be expressed in the following form:
\begin{align}
  x'_{ij} = (1 + \lambda_i \sigma_{ij}) x_{ij}, ~ \text{ with } ~ -\kappa \leq \lambda_i \leq \kappa
\end{align}

They took $min$ of $\sigma_{ij}^{(x)}$ and $\sigma_{ij}^{(y)}$ so that pixels along coordinate-aligned edges are not changed.

\section{Localized (Patch) attack}

Instead of perturbing the whole image, some attack methods try to perturb only a localized region in the image, or sometimes referred to as ``patch attack'' because the perturbed region resembles a patch attached to the image.

Karmon et al. \cite{karmon2018lavan} created localized and visible adversarial noise (LaVAN) that cover only 2\%
of the pixels in the image without covering the
main object. In this method, visible noise is added to a local position of the image to produce an adversarial example. To confine the noise $\delta$ to a small area over
the image $x$, they used a mask $m \in {\{0, 1\}}^n$, and the noised image is defined as $(1 - m) \odot x + m \odot \delta$, where $\odot$ is element-wise multiplication. It is worth noticing that the noise is used to replace the area rather than be added to it. As the term ``visible'' suggests, the adversarial examples generated by this method are aimed to be easily detected by human.

Dia et al. \cite{dia2021localized} proposed localized uncertainty attacks, which is a novel class of adversarial attacks that creates adversarial examples against deep learning models through replacement of \textit{uncertain} regions in the original inputs. Instead of perturbing inputs indiscriminately, they utilize the uncertainty associated with the predictions of the classifier and focus only on carefully chosen regions, to yield better imperceptibility. They proposed to use a binary-valued (with values in $\{0, 1\}$) mask $\omega$ with the same size of the input $x$, and apply the perturbation to the region indicated by $\omega \odot x$ to generate adversarial examples. They trained a feed-forward network called \textit{mask model} to learn a distribution $p_v(x)$ over all possible masks $\omega$ that, when drawing a mask from $\omega \sim p_v(x)$ and applying the perturbation to the region $\omega \odot x$, the predictive uncertainty of the classifier is maximized. Ideally, masks that lead to greater uncertainty should have a higher probability in the distribution $p_v(x)$.

Bai et al. \cite{bai2021inconspicuous} proposed Inconspicuous Adversarial Patch Attack (IAPA), which generates inconspicuous adversarial patches using GANs. Compared with other adversarial patches attacks, IAPA causes less modifications, and has higher chance to evade detection from human. In their approach, they first used a GradCAM  \cite{selvaraju2016grad} based ``vulnerability map'' to decide perceptual sensitivity of the victim model. Based on the vulnerability map $M$, they then choose the area with the highest importance as the victim area for patch attack. However, since GradCAM can only work for Convolutional Neural
Network (CNN) based models, their approach can not be applied to Non-CNN based models.

\section{Summary}

We introduced DeepFool attack and Brendel\&Bethge (BB) attack in this chapter, providing methods of finding adversarial examples and minimizing the $L_p$-norm of the adversarial perturbation, which are adopted in our system and integrated with mask-constraint.

Two works on imperceptible attacks are introduced, which showed that variance of image provides useful guidance on generating imperceptible adversarial perturbations. However, they both use heuristics to select pixels and apply perturbation, so they have no control on which pixels are selected. Also, they didn't give sufficient results to quantitatively evaluate the imperceptibility of their attack.

Three Localized (Patch) attacks are introduced, but they have different problems that make them not suitable for our need. LaVAN generate adversarial examples easy to be detected by human, localized uncertainty attack neither have control on the attack region nor consider imperceptibility, IAPA suggests a good way to find vulnerable regions, but only applies to convolutional models.

\SetPicSubDir{ch-Method}
\SetExpSubDir{ch-Method}

\chapter{\attackname{}}
\label{ch:method}
\vspace{2em}

In this section, we introduce the framework and formulation of our proposed adversarial perturbation generation system named \attackname{}. The complete workflow of our system is shown in \autoref{method:fig:workflow}. The system requires a clean image $x$ and a Deep Learning classifier $f$. A mask generator is first used to generate a mask-constraint $\mathcal{E}$. The generator takes the clean image $x$ and a series of parameterized user specification as inputs, and either generate the mask-constraint automatically or prompt the user to specify it through a GUI. Details of the mask generator interface can be found in \autoref{sec:meth:GUI} and \autoref{sec:meth:system}. A preliminary adversarial example $x'$ is then generated by using DeepFool attack under the mask-constraint $\mathcal{E}$. Finally, Brendel\&Bethge (BB) attack is used to minimize the $L_p$-norm distance between $x'$ and $x$ and get the optimized adversarial example $x''$ under the same constraint $\mathcal{E}$.
\begin{figure}[h]
  \centering
  \includegraphics[width=.8\linewidth]{\Pic{png}{workflow}}
  \vspace{\BeforeCaptionVSpace}
  \caption{Workflow of our proposed \attackname{}.}
  \label{method:fig:workflow}
\end{figure}

\section{The Basic Form: $L_\infty$ Attack}

Our system performs $L_\infty$ attack based on the algorithms of DeepFool and Brendel \& Bethge (BB) attacks, as we believe these two algorithms are good complement for each other based on our experiment results in \autoref{sec:eval:imperc}. DeepFool $L_\infty$ attack finds adversarial example near the decision boundary but does not guarantee to have minimal $L_\infty$ distance from the clean image. BB $L_\infty$ attack finds adversarial example on the decision boundary with minimized $L_\infty$ distance from the clean image, but relies heavily on the quality of the starting point. The attack method we use in our system is a combination of these two brilliant methods, as shown in \autoref{Method:algo:1}. 

In summary, our system finds an adversarial example on the decision boundary with minimal $L_\infty$ distance from the clean image, by first searching for a small adversarial example close to the decision boundary (\autoref{algo:startDF} to \autoref{algo:endDF} in \autoref{Method:algo:1}). It then minimizes the $L_\infty$ distance from the clean image while the subject image remains an adversarial example (\autoref{algo:startBB} to \autoref{algo:endBB} in \autoref{Method:algo:1}). Different from the original methods where the constraint on each pixel is homogeneous i.e. each adversarial example should stay within the $L_\infty$ ball $\{x + \delta ~|~ \vbarbar{\delta}_\infty \leq \epsilon\}$, we use a general mask-constraint as defined in \autoref{def:mask} so that the adversarial examples are constrained in $\{x + \delta ~|~ \forall_{\delta_i \in \delta} ~ \vbar{\delta_i} \leq \epsilon_i\}$. 

\begin{algorithm}[!t]
\AlgoFontSize
\DontPrintSemicolon

\KwGlobal{loosen rate of the constraint $\lambda$}
\KwGlobal{per \# of iteration to loosen the constraint $T$}
\BlankLine

\SetKwFunction{fAttack}{Attack}
\SetKwFunction{fOptimize}{Optimize}

\KwIn{clean image $x$, classifier $f$, mask constraint $\mathcal{E}$, maximum iterations $max\_iter$}
\KwOut{adversarial image $x'$}
\Proc{\fAttack{$n$}}{
  Initialize $x^0 \gets x, i \gets 0$\; \label{algo:startDF}
  \While{$\hat{y}(x^i) = \hat{y}(x^0)$ and $i < max\_iter$}{
    \For{$k \neq \hat{y}(x^0)$}{
      $f'_k \gets f_k(x^i) - f_{\hat{y}(x^0)}(x^i)$\;
      $\omega'_k \gets \nabla f'_k$\;
    }
    $\hat{l} \gets \argmin_{k \neq \hat{y}(x^0)}{\frac{\vbar{f'_k}}{\vbarbar{\omega'_k}_1}}$\;
    $\delta^i \gets \frac{\vbar{f'_{\hat{l}}}}{\vbarbar{\omega'_{\hat{l}}}_1}sign(\omega'_{\hat{l}})$\;
    \If{$i > 0$ and $i \mod T = 0$}{\label{algo:updateC_start}
      \For{$\epsilon$ in $\mathcal{E}$}{
        $\epsilon \gets \epsilon * \lambda$\;
      }
    }\label{algo:updateC_end}
    $x^{i+1} \gets clip(x^i + \delta^i, \mathcal{E})$\;
    $i \gets i + 1$\;
  }
  $x' \gets x^i$\; \label{algo:endDF}
  \If{$\hat{y}(x') = \hat{y}(x^0)$}{
    \Return $None$ \tcp{Attack Failed}
  }
  \Else{
    $x' \gets \fOptimize{x, f, x', C}$\;
    \Return $x'$\;
  }
}

\BlankLine

\KwIn{clean image $x$, classifier $f$, starting point $\tilde{x}^0$, mask constraint $\mathcal{E}$}
\KwOut{optimized adversarial image $x''$}
\Func{\fOptimize{$x, f, \tilde{x}^0, C$}}{
  Initialize $i \gets 0, b^0 \gets 0$\; \label{algo:startBB}
  \While{$i < $maximum number of steps}{
    $c^i \gets \min_{k, k \neq \hat{y}(x)}{(f_{\hat{y}(x)}(\tilde{x}^{i-1}) - f_k(\tilde{x}^{i-1}))}$\;
    $b^i \gets \nabla_{\tilde{x}^{i-1}}c^i$\;
    $\delta^i \gets$ Solve \autoref{eq:BB} under constraint C for $L_\infty$ norm\;
    $\tilde{x}^i \gets \tilde{x}^{i-1} + \delta^i$\;
    $i \gets i + 1$\;
  }
  $x'' \gets x^i$\; \label{algo:endBB}
  \Return $x''$\;
}

\caption{Attack procedure of \attackname{}.}
\label{Method:algo:1}
\end{algorithm}

\section{With Focus: Regional Attack}
\label{sec:meth:GUI}
One major advantage of our system is the flexibility to add perturbation to any sub-region of the clean image, making our system able to not only simulate whole-image attacks, but also simulate regional attacks that are more practical in the physical world. 

In order to allow users to define their own mask-constraint in a convenient and unified way, we implemented a Graphical User Interface to enable users to indicate regions on which they want to focus. As shown in \autoref{method:fig:gui1} and \autoref{method:fig:gui2}, a user can specify the target region for an attack either by clicking and dragging on the screen to select a rectangle shaped region, or by clicking and brushing to select an arbitrary shaped region.

\begin{figure}[t]
  \centering
  \begin{minipage}[t]{.45\linewidth}
    \centering
    \includegraphics[width=\linewidth]{\Pic{png}{rect_mask}}
    \caption{Rectangle Shaped Region.}
    \label{method:fig:gui1}
  \end{minipage}
  \hspace*{2em}
  \begin{minipage}[t]{.45\linewidth}
    \centering
    \includegraphics[width=\linewidth]{\Pic{png}{brush_mask}}
    \caption{Arbitrary Shaped Region.}
    \label{method:fig:gui2}
  \end{minipage}
\end{figure}

\section{Imperceptible Attack: Variance Map}
\label{sec:meth:imperc}
Images with small $L_\infty$ distances do not necessarily resemble one another. Following the same idea of imperceptible attacks \cite{luo2018towards,croce2019sparse} introduced in \autoref{sec:review:imperc}, our system is also capable of generating imperceptible adversarial perturbations. This can be easily achieved with the mask-constraint of our system. For our imperceptible attack, we adapted the variance map proposed by Croce et al. \cite{croce2019sparse}, as in \autoref{eq:review:imper:4}. 

\subsection{Non-adaptive Imperceptible Attack}
\label{ssec:non-adap}
We first introduce our Non-adaptive Imperceptible Attack with fixed mask-constraint. This attack is considered ``non-adaptive'' as it does not take into consideration the robustness properties of the subject classifier.

Given an image $x \in \mathbb{R}^{d \times c}$ with $d$ pixels and $c$ channels, where $x_{ij}$ represents the value of the $i_{th}$ pixel in the $j_{th}$ channel. The variance of $x_{ij}$ in the variance map is calculated as:
\begin{align}
  \sigma_{ij} = min(\sigma_{ij}^{(x)}, \sigma_{ij}^{(y)})
\end{align}
where $\sigma_{ij}^{(x)}$, $\sigma_{ij}^{(y)}$ are the standard deviation of each color channel in x- and y-axis directions with the two immediate neighboring pixels and the original pixel $x_{ij}$.

We use the variance map to generate a mask-constraint $\mathcal{E}$: For each $x_{ij}$, the mask-constraint $\epsilon_{ij}$ on it is defined as $\epsilon_{ij}=\sigma_{ij}$, so that the perturbation $\delta_{ij}$ satisfies $-\sigma_{ij} \leq \delta_{ij} \leq \sigma_{ij}$.

The adversarial example $x'$ with imperceptible perturbation is then obtained by applying our $L_\infty$ attack on image $x$ with the mask-constraint $\mathcal{E}$.

\subsection{Adaptive Imperceptible Attack}

The variance map as computed in \autoref{ssec:non-adap} is purely based on the input image, without considering the robustness properties of the classifier; thus rendering it futile in providing attack for many robust models. To circumvent this shortcoming, Croce et al. \cite{croce2019sparse} used a hyper-parameter $\kappa$ to tune the constraints to handle models with different robustness; their method however requires users to perform trial-and-error on $\kappa$ value under their experiment settings, and does not consider the variation of the model's robustness on different inputs. 

In our work, we improve over their static variance map method by removing the hyper-parameter $\kappa$, and using an adaptive method to automatically loosen the constraint for models with different robustness on different inputs, as shown in \autoref{algo:updateC_start} to \autoref{algo:updateC_end} of \autoref{Method:algo:1}. We set the loosen rate $\lambda$ of the constraint and the interval $T$ (number of iterations) between each loosening of the constraint. 

When running DeepFool attack, we start with the initial mask-constraint $\mathcal{E}$ as described in \autoref{ssec:non-adap}, and multiple each constraint $\epsilon \in \mathcal{E}$ by $\lambda$ at the end of each $T$ iteration.
\begin{align}
  \epsilon \gets \epsilon * \lambda\: \text{ for all } \epsilon \in \mathcal{E}
\end{align}

With the steady loosening of the constraint, our adaptive imperceptible attack can find good $\kappa$ values for different inputs by itself, and obtain higher attack success rate than the non-adaptive imperceptible attack in \autoref{ssec:non-adap}.

\section{Vulnerability Estimation: Importance Map}
\label{sec:meth:regional}
While it is possible to apply perturbation to any sub-region of the clean image in our system, chances are that no adversarial examples can be found under such regional constraints, or the resulting adversarial perturbation may require too large $L_\infty$-norm value that exceeds our expectation. It's difficult and yet interesting to know which pixels/sub-regions are the most vulnerable to adversarial attacks, i.e., having the smallest robust radius. Formally, We define the robust radius of a region as:

\begin{defn}[\textbf{Robust Radius of a Region}]
Given an image classification neural network $N$ with prediction function $\hat{y}(x)$, an image $x = (x_1, x_2, ..., x_d)$, a region specified by a binary mask $\omega$ with binary values 0's and 1's and has the same size with $x$. We define the robust radius of the region $\omega$ to be:
\begin{align}
  \begin{split}
  &r_*(\omega)= min(r) \\
  &\text{ s.th. } \exists x'\: \hat{y}(x') \neq \hat{y}(x) \: \text{and} \\ &\vbar{x'_i - x_i}
  \begin{cases}
    =0,          & \text{if } \omega_i=0\\
    \leq r,    & \text{if } \omega_i=1
  \end{cases} \text{for i = 1, ..., d}
  \end{split} \label{eq:robustregion}
\end{align}
\end{defn}

In this section, we investigate the problem of finding the most vulnerable region:

\begin{problem}[\textbf{Most vulnerable region}]
Given a image classification neural network $N$ with prediction function $\hat{y}(x)$, an image $x = (x_1, x_2, ..., x_d)$, a set of regions specified by binary masks $\Omega$, where each $\omega \in \Omega$ is a mask with the same size as $x$ and has binary values 0's and 1's component-wise. Determine the region most vulnerable to adversarial attack, that is:
\begin{align}
\argmin_{\omega \in \Omega}r_*(\omega) \label{eq:vulnerable}
\end{align}
\end{problem}

Although it's difficult to solve the exact problem above, our system estimates the robust radius (hence the vulnerability) of a region by running $L_\infty$ attack on the region. However, when the number of regions under consideration becomes large, running $L_\infty$ attack on each region can be time-consuming with finite computing power.

We therefore propose to use an importance map to capture the importance of each pixel on affecting the model's prediction, and to offer insight on how to select a ``good'' sub-region with high vulnerability. Different from Bai et al. \cite{bai2021inconspicuous} who used GradCAM to capture the pixel importance for only CNN models, we use Integrated Gradients attribution algorithm \cite{sundararajan2017axiomatic} with SmoothGrad \cite{smilkov2017smoothgrad} to generate the importance map, which is applicable to any deep learning models.

Given a clean image $x = (x_1, x_2, ..., x_d)$ and a classifier $f$, we use Integrated Gradients with SmoothGrad to calculate an importance value $I_i$ for each pixel $x_i$, and the values for all pixels forms an importance map $I$ with the same size of $x$. We then use the importance map to estimate the vulnerability of a region (specified by a binary mask $\omega$) by summing up the importance value of all pixels in that region, as shown in \autoref{eq:imp_sum}.
\begin{align}
  vulnerability(\omega) =& \sum_1^d I_i * \omega_i
  \label{eq:imp_sum}
\end{align}
The most vulnerable region is then determined by:
\begin{align}
  \argmax_{\omega \in \Omega} \sum_1^d I_i * \omega_i
  \label{eq:mvr}
\end{align}

Intuitively, as higher importance values tend to change the classifier's decision more significantly, they can attain higher vulnerability. This intuition is supported by our empirical study in Section \autoref{sec:eval:vulnerability}.

\subsection{Integrated Gradients}

In simple terms, Integrated Gradients define the attribution (importance) of the input's $i^{th}$ feature as the path integral of the line path from a baseline $x^b_i$ to the input $x_i$, as shown in \autoref{eq:IG}
\begin{align}
  \begin{split}
  IntegratedGradients_i(x) = (x_i - x^b_i) \times \int_{\alpha=0}^{1} \pdv{F(x^b+\alpha(x-x^b))}{x_i}d\alpha
  \end{split} \label{eq:IG}
\end{align}
where $F$ represents a neural network, $\pdv{F(x))}{x_i}$ is the gradient of $F(x)$ on the $i^{th}$ dimension. It is recommended to use baseline $x^b$ that brings network output close to zero, e.g. an all black image. In practice, the integral can be approximated by interpolating the linear path from $x^b_i$ to $x_i$ and then sum the gradients of those interpolations, as in \autoref{eq:IG_approx}
\begin{align}
  \begin{split}
  IntegratedGradients^{approx}_i(x) = (x_i - x^b_i) \times \sum_{k=1}^{m} \pdv{F(x^b+\frac{k}{m}(x-x^b))}{x_i} \times \frac{1}{m}
  \end{split} \label{eq:IG_approx}
\end{align}

\subsection{SmoothGrad}

\begin{figure}[h]
  \centering
  \includegraphics[width=.6\linewidth]{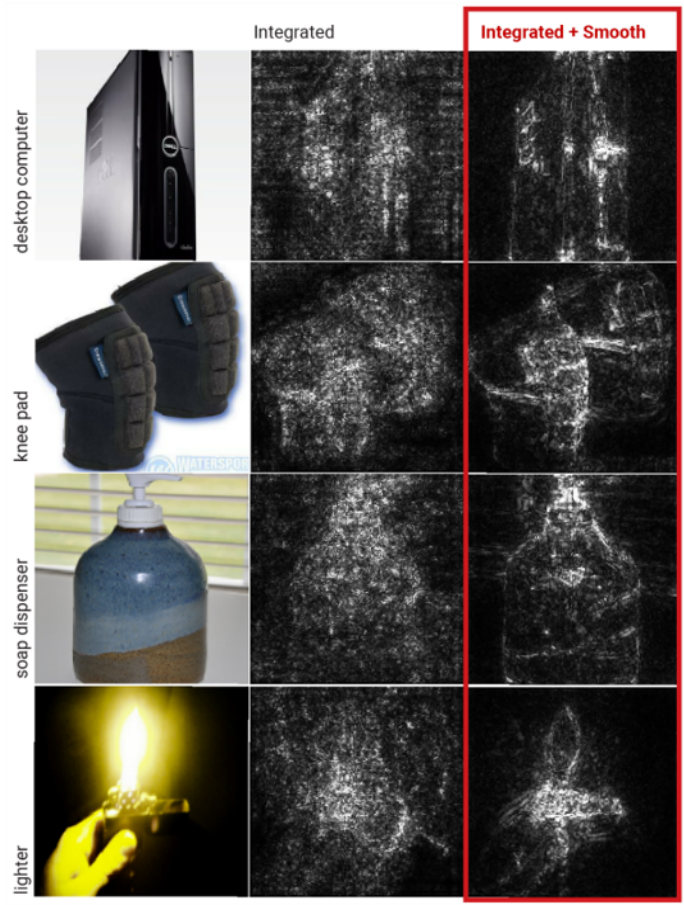}
  \vspace{\BeforeCaptionVSpace}
  \caption{Using SmoothGrad to remove noise in importance maps generated by Integrated Gradients \cite{smilkov2017smoothgrad}.}
  \label{method:fig:SmoothGrad}
\end{figure}

The gradient of $F$, however, may fluctuate sharply at small scales, causing the importance maps to be visually noisy. As point out in \cite{smilkov2017smoothgrad}, given such sharp fluctuations, the gradient of $F$ at any given point is less meaningful than the local average of gradient values.
In replacement of the gradient of $F$, SmoothGrad removes the noise in importance maps, as shown in \autoref{method:fig:SmoothGrad}, by taking random samples in a neighborhood of the input $x$ and averaging the results, as expressed in \autoref{eq:SmoothGrad}.
\begin{align}
  \hat{I} (x) = \frac{1}{n}\sum_{1}^{n} I (x + \mathcal{N}(0, \sigma^2))
  \label{eq:SmoothGrad}
\end{align}
where $n$ is the number of samples, and $\mathcal{N}(0, \sigma^2)$ represents Gaussian noise with standard deviation $\sigma$.

\subsection{Correction Coefficient}

While the importance map $I$ can be obtained using Integrated Gradients with SmoothGrad method, naive perturbations of sub-regions with higher importance do not necessarily produce adversarial images with smaller $L_\infty$ distance to the clean image. For example, a black car in the image might be a strong sign that the image should be classified as ``car'', hence the black car region (with pixel value $\approx$ 0) could have high importance. But when adding perturbation to the black area, many of the pixels could be clipped at 0 to keep the image valid, which would cause the other pixels to be perturbed more than expected and result in large $L_\infty$ distance. To solve this problem, we proposed to add a correction coefficient $\beta$ to the importance map, as shown in \autoref{eq:correction}.
\begin{align}
  \begin{split}
  \beta_i = 0.5 + min(x_i, 1-x_i)
  \end{split} \label{eq:correction}
\end{align}
where the pixel with a value of 0.5 has a coefficient of 1, and the pixel with a value of 0 (black) or 1 (white) has a coefficient of 0.5, as we assume they have a 50\% chance of contributing to the perturbation without being clipped.

After adding the correction coefficient $\beta$ to the importance map $I$, the vulnerability of a region $\omega$ defined in \autoref{eq:imp_sum} now becomes:
\begin{align}
  vulnerability(\omega) =& \sum_1^d \beta_i * I_i * \omega_i
  \label{eq:imp_sum_c}
\end{align}
and the most vulnerable region defined in \autoref{eq:mvr} becomes:
\begin{align}
  \argmax_{\omega \in \Omega} \sum_1^d \beta_i * I_i * \omega_i
  \label{eq:mvr_c}
\end{align}

\subsection{Efficiency of the method}

The set of all possible regions can be excessively large, for example, the set of all $h \times w$ rectangle sub-regions in a image with size $H \times W$ has a size of $(H-h+1) \times (W-w+1)$. To have a good estimation of the vulnerability of each region and select the most vulnerable region, we need to run our $L_\infty$ attack for $(H-h+1) \times (W-w+1)$ times. For example, to find the most vulnerable $10 \times 10$ region in a $28 \times 28$ image (which is small), we need to run $L_\infty$ attack for $19 \times 19 = 361$ times. As running $L_\infty$ attack on a single region can cost a few seconds on some large networks, the time cost can easily reach thousands of seconds. 

To avoid this enumerative search, we propose to use the importance map base method to find the most vulnerable region according to \autoref{eq:rect_region}:
\begin{align}
  \begin{split}
  \argmax_{x \leq W-w+1, y \leq H-h+1} \sum_{i=x}^{x+w}\sum_{j=y}^{y+h}\beta_{ij}I_{ij}
  \end{split} \label{eq:rect_region}
\end{align}
where $(x, y)$ is the index of the top-left corner of the sub-region, $I_{ij}$ is the value of pixel $(i, j)$ in the importance map, $\beta_{ij}$ is the correction coefficient  for pixel $(i, j)$. Instead of running $L_\infty$ attack on each region, this method only calculates the sum of importance score for each region and doesn't involve any execution of the $L_\infty$ attack, which makes it a lot faster and scale to large number of regions.

We show later in the experiments that our importance map is relatively good in measuring vulnerability, so, when given a set of regions for consideration, our importance map can be used to select the most vulnerable region to apply perturbation in order to have a higher chance of success and smaller $L_\infty$-norm of the perturbation.

It is important to note that the importance map based method is based on heuristics and therefore may not estimate the vulnerability as good as the $L_\infty$ attack based method. Therefore, we also propose a way to extend it to estimate vulnerability more accurately, by broadening our range of selection to consider the k (instead of 1) candidate regions with the highest importance scores among all regions, and run $L_\infty$ attack to calculate the vulnerability of each region, then select the region that is actually the most vulnerable. This extended method sacrifices time efficiency for better estimation accuracy.

\section{The System as a Whole}
\label{sec:meth:system}
With all the aforementioned techniques integrated in our system, it provides much flexibility for users to explore various kinds of adversarial examples as desired. In the basic form, our system is an $L_\infty$ attack system that combines two existing attack methods: DeepFool and Brendel\&Bethge attack. The system further allows users to indicate regions of their focus through a convenient Graphical User Interface, and to specify whether they want the attack to be imperceptible, and how many pixels they want to perturb.

We implemented a Command Line interface for users to specify the property of the attack. The arguments are listed and explained below:
\begin{description}

\item[eps]: Float. Limitation of the perturbation magnitude on each pixel.

\item[region]: \{`whole', `select'\}. The region to apply perturbation. `whole' means the perturbation is on the whole image, whereas `select' invokes the GUI and prompts users to select the region.

\item[imperceptible]: Boolean. Whether to produce an imperceptible attack.

\item[ratio]: Float. When 0 < ratio < 1, it represents the maximum percentage of the pixels allowed to be perturbed. When ratio > 1, it represents the number of pixels allowed to be perturbed.

\end{description}

\section{Summary}

In this chapter, we introduced our system in detail. We adopted and combined two existing attack methods -- DeepFool and Brendel\&Bethge attack. We integrated them with our mask-constraint. We proposed to use variance-based mask-constraint to generate imperceptible adversarial perturbations and adaptively loosen the constraint to handle more robust models. We formulated the problem of finding a vulnerable region within an image and proposed an importance map method to approximately but efficiently solve this problem. We proposed to use Integrated Gradients with SmoothGrad to generate importance maps; the proposed method is applicable to any deep learning models. Finally, we further proposed a correction coefficient to fix a technical glitch in using importance maps for vulnerability estimation.
\SetPicSubDir{ch-Evaluation}
\SetExpSubDir{ch-Evaluation}

\chapter{Experiments \& Results}
\label{ch:evaluation}
\vspace{2em}

To access the effectiveness of \attackname{}, we conducted a series of experiments using several neural networks for image classification on different datasets. We also conduct user study to evaluate users’ perceived satisfaction of our system.

\section{Experimental Setup}
\label{sec:eval:setup}

\textbf{Datasets.} We consider three commonly-used datasets for image classification and adversarial robustness benchmarking, including MNIST \cite{lecun-mnisthandwrittendigit-2010}, CIFAR10 \cite{krizhevsky2009learning} and STL10 \cite{coates2011analysis}.

\textbf{Models.} We select a variety of models to thoroughly evaluate attacks under different conditions. Most of the models are pre-trained models obtained from ERAN's github page\footnote{https://github.com/eth-sri/eran}. For MNIST, we consider the following two models: M1, a 9-layer undefended fully-connected network with 1610 units and ReLU activation; M2, a 3-layer undefended convolutional network with 3,604 units and ReLU activation. For CIFAR10, we consider four models: C1, a 6-layer undefended fully-connected network with 3000 units and ReLU activation; C2, a 3-layer undefended convolutional network with 7,144 units and Sigmoid activation; C3, a 6-layer convolutional network with 62,464 units and ReLU activation, adversarial trained using DiffAI \cite{mirman2018differentiable}; and C4, a 19-layer residual network with 558K units and ReLU activation, adversarial trained using PGD \cite{madry2017towards}. For STL10, we used the pre-trained 6-layer convolutional network S1 from this github repository\footnote{https://github.com/aaron-xichen/pytorch-playground}, which takes normalized inputs with values ranging from -1 to 1. The models are listed in \autoref{Tab:models}.

\begin{table}[]
\centering
\resizebox{\columnwidth}{!}{
\begin{tabular}{|c|c|c|c|c|c|c|c|c|}
\hline
dataset & image size & model & architecture & \#layers & \#units & activation & training defense & accuracy \\ \hline
\multirow{2}{*}{MNIST} & \multirow{2}{*}{28*28} & M1 & fully connected & 9 & 1610 & ReLU & None & 0.95 \\ \cline{3-9} 
 & & M2 & convolutional & 3 & 3604 & ReLU & None & 0.98 \\ \hline
\multirow{4}{*}{CIFAR10} & \multirow{4}{*}{32*32} & C1 & fully connected & 6 & 3000 & ReLU & None & 0.56 \\ \cline{3-9} 
 & & C2 & convolutional & 3 & 5704 & Sigmoid & None & 0.55 \\ \cline{3-9} 
 & & C3 & convolutional & 6 & 48064 & ReLU & DiffAI & 0.51 \\ \cline{3-9} 
 & & C4 & residual & 19 & 558K & ReLU & PGD & 0.82 \\ \hline
STL10 & 96*96 & S1 & convolutional & 5 & 652K & ReLU & None & 0.77 \\ \hline
\end{tabular}
}
\caption{Datasets and Models.}
\label{Tab:models}
\end{table}

\textbf{Attacks.} For $L_\infty$ attack, we compare our method against different state-of-the-art attacks for finding adversarial perturbations with minimum $L_\infty$ norm: the DeepFool attack \cite{moosavi2016deepfool}, the Brendel \& Bethge (BB) attack \cite{brendel2019accurate}, the Fast Adaptive Boundary (FAB) attack \cite{croce2020minimally}, and the Fast Minimum-norm (FMN) attack \cite{pintor2021fast}. We used the open-source implementations of DeepFool and FAB from AdverTorch\footnote{https://github.com/BorealisAI/advertorch}, and implementations of BB and FMN from Foolbox\footnote{https://github.com/bethgelab/foolbox}. In the other experiments, we only use the attack method of our system.

\textbf{Hyper-parameters.} To ensure a fair comparison, we used similar default settings for each attack. We report here the hyper-parameters used in each experiments. For $L_\infty$ attack, we configured each attack to have 100\% Attack Success Rate for all models and all datasets, we set the max allowable perturbation size $\epsilon = 1.0$ for all attacks, the hyper-parameter configurations for each attack are detailed below.

\textit{DeepFool.} We set the max number of iterations to be 50, and overshoot be 0.02.

\textit{FAB.} We set the max number of iterations to be 100, bound of step bias $\alpha_{max} = 0.1$, extrapolation step $\eta=1.05$, backward step $\beta=0.9$, with no random restarts.

\textit{BB.} We used the default initial attack of BB, called Random-Noise attack, which randomly draws 1,000 directions to search for adversarial examples and takes 1,000 binary search steps to blend adversarial example and original image in each direction. The adversarial example with the smallest $L_p$ distance to the original image is selected as the starting point. We set the number of iterations to be 100, number of binary search steps to be 10, the trust radius decays every 20 iterations with a coefficient of 0.5.

\textit{FMN.} We set the number of iterations to be 100, number of binary search steps to be 10, the decaying step size $\gamma$ starting from 0.05.

\textit{\attackname{}} For DeepFool, we set the max number of iterations to be 50, and overshoot be 0.02. For BB, we set the number of iterations to be 100, number of binary search steps to be 10, the trust radius decays every 20 iterations with a coefficient of 0.5.

For Imperceptible attack, we used the $L_\infty$ attack of \attackname{} with the same configuration described above. For adaptive loosening of the constraint, we set the loosen\_rate to be 1.2, and loosen the constraint every 10 iterations.

For Vulnerable Region estimation, we also used the $L_\infty$ attack of \attackname{} with the same configuration described above to estimate the robust radius.

\textbf{Metrics.} For $L_\infty$ attack, we report the average $L_\infty$ norm of the perturbations. For Imperceptible attack, we report the Attack Success Rate (ASR) for each attack, the $L_0$, $L_2$ and $L_\infty$ norm of the perturbations, the structural similarity (SSIM) between original images and adversarial examples, and perceptual color distance (CIEDE2000) between original images and adversarial examples for colored datasets (CIFAR10, STL10). For Vulnerable Region estimation, we report the minimal robust radius among all regions for an image and average it over 100 images, we also report the robust radius of the regions found by different importance map based method. To measure users' satisfaction of our system, we issued and collected some PSSUQs (Post-Study System Usability Questionnaire) \cite{lewis1992psychometric} to measure the System Usefulness (SYSUSE), Information Quality (INFOQUAL) and Interface Quality (INTERQUAL) of our system.

\section{Evaluation of $L_\infty$ attack}
\label{sec:eval:linf}
To evaluate the performance of \attackname{} in $L_\infty$ attack, we first compare it to state-of-the-art $L_\infty$ adversarial attacks. We calculate over the first 100 correctly classified images for each dataset
%
, the average $L_\infty$ norm of the adversarial perturbations generated by each attack on the whole image, where smaller average $L_\infty$ norm indicates better performance, as shown in \autoref{Tab:linf}. FAB is not evaluated for S1 because their official implementation does not support input values ranging from -1 to 1. 

\begin{table}[h]
\centering
\begin{tabular}{|c|c|c|c|c|c|}
\hline
& DEEPFOOL & FAB & BB & FMN & \attackname{} \\ \hline
M1 & 0.07423 & 0.06379 & 0.06358 & \textbf{0.06153} & 0.06182 \\ \hline
M2 & 0.1837 & 0.1571 & 0.1558 & 0.1692 & \textbf{0.1533} \\ \hline
C1 & 0.01328 & 0.008917 & 0.009386 & 0.008186 & \textbf{0.008179} \\ \hline
C2 & 0.01135 & 0.01082 & 0.01236 & 0.01169 & \textbf{0.01074} \\ \hline
C3 & 0.01731 & 0.01433 & 0.01424 & 0.01397 & \textbf{0.01370} \\ \hline
C4 & 0.04297 & 0.03434 & 0.03629 & 0.03477 & \textbf{0.03368} \\ \hline
S1 & 0.005805 & - &  0.004537 & 0.004575 & \textbf{0.004366} \\ \hline
\end{tabular}
\caption{Average $L_\infty$ norm for different attacks} 
\label{Tab:linf}
\end{table}

The results show that by combining the strength of DeepFool and Brendel \& Bethge attack, \attackname{} achieves comparable results to the state of the art methods on different datasets, model architectures and training methods.

We also compare the execution time taken by each attack. \autoref{Tab:linf_time} shows the average execution time on one image for different attacks, the experiment was done on the ResNet18 model $C4$ on CIFAR10 dataset. \attackname{} took moderate execution time among all the attacks, which is acceptable considering its specific objective in generating minimal $L_\infty$ perturbations.

\begin{table}[h]
\centering
\begin{tabular}{|c|c|c|c|c|c|}
\hline
& DEEPFOOL & FAB & BB & FMN & \attackname{} \\ \hline
Time & 1.97 & 29.43 & 19.22 & 2.96 & 7.62 \\ \hline
\end{tabular}
\caption{Average execution time (seconds) for different attacks} 
\label{Tab:linf_time}
\end{table}

\subsection{Discussion on the result of BB and \attackname{}}
Although \attackname{} uses the same method for minimizing the $L_\infty$-norm of the perturbation as that of BB, it outperforms BB by having shorter execution time, which we attribute to the quality of the starting point and the method to find it.

As mentioned in \autoref{sec:eval:setup}, by default, BB use a Random-Noise attack, BB will randomly draw 1,000 directions to search for adversarial examples and take 1,000 binary search steps to blend adversarial example and original image in each direction. In each step, it needs to run a forward pass of neural network, resulting in 1,000,000 forward passes in total, while DeepFool only runs at most 50 forward passes, which makes it a lot faster.

The Random-Noise attack used by BB is also not efficient, as the number of all possible directions is $2^N$, where N is the number of pixels in the image. The search space is so large that it is infeasible to cover the optimal one in 1,000 randomly drawn directions. On the contrary, DeepFool utilizes local gradient information to search for only the most promising direction.

\section{Evaluation of Variance Map based Imperceptible attack}
\label{sec:eval:imperc}
We evaluate here the effectiveness of our variance map based imperceptible attack by comparing it with our $L_\infty$ attack which represents normal $L_p$-norm based attacks. We assess the imperceptibility of an adversarial attack according to the SSIM \cite{wang2004image} and CIEDE2000 \cite{luo2001development} measures between the adversarial image generated by the attack and the original image, which are mentioned in \autoref{sec:intro:limitation}. We also provide image examples for human assessor to assess the effectiveness of our imperceptible attack.

\begin{figure}[h]
  \centering
  \includegraphics[width=.8\linewidth]{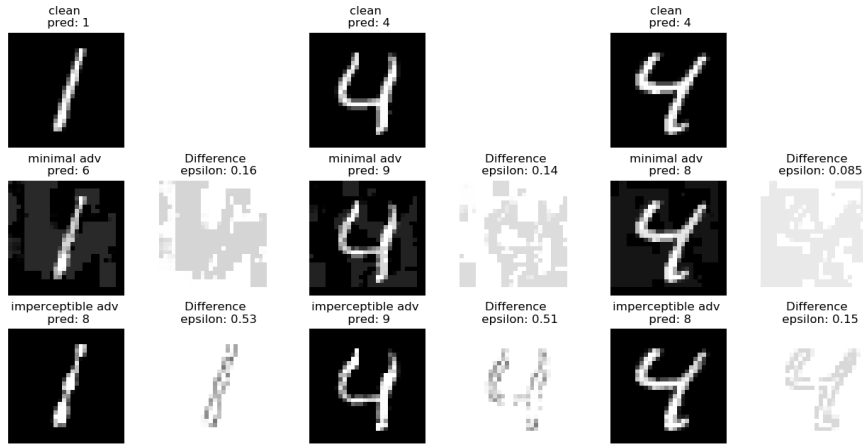}
  \vspace{\BeforeCaptionVSpace}
  \caption[$L_\infty$ and Imperceptible attack on MNIST.]{\textbf{$L_\infty$ and Imperceptible attack on MNIST.} We illustrate the differences of the adversarial examples found by $L_\infty$ attack and Imperceptible attack. From top to bottom in each group. \textit{top row} - original image, \textit{middle row} - adversarial examples found by $L_\infty$ attack, and the adversarial perturbations, \textit{bottom row} - adversarial examples found by Imperceptible attack, and the adversarial perturbations.}
  \label{eval:fig:imperc1}
\end{figure}

\begin{figure}[h]
  \centering
  \includegraphics[width=.8\linewidth]{\Pic{png}{cifar10_imperc}}
  \vspace{\BeforeCaptionVSpace}
  \caption[$L_\infty$ and Imperceptible attack on CIFAR10.]{\textbf{$L_\infty$ and Imperceptible attack on CIFAR10.} We illustrate the differences of the adversarial examples found by $L_\infty$ attack and Imperceptible attack. From top to bottom in each group. \textit{top row} - original image, \textit{middle row} - adversarial examples found by $L_\infty$ attack, and the adversarial perturbations, \textit{bottom row} - adversarial examples found by Imperceptible attack, and the adversarial perturbations.}
  \label{eval:fig:imperc2}
\end{figure}

We first illustrate the differences between adversarial examples found by our $L_\infty$ and imperceptible attacks. This serves to dispel a common misconception that adversarial attack images is naturally imperceptible, and vice versa.  The imperceptible attack we use here is the adaptive version introduced in \autoref{sec:meth:imperc}. Some examples are shown in \autoref{eval:fig:imperc1} and \autoref{eval:fig:imperc2}. The adversarial examples found by $L_\infty$ attack, while have small $L_\infty$ distance (epsilon) to the original images, do not resemble the original images because of the noise introduced by the perturbation at low variance regions. The imperceptible attack that is constrained by the variance map, generates adversarial examples that are more similar to the original images according to human perception. 

We further quantitatively evaluate the effective of our imperceptible attack, over 100 correctly classified image samples on each dataset, as shown in \autoref{Tab:imperc}. We compare the unconstrained $L_\infty$ attack with the non-adaptive imperceptible attack and adaptive imperceptible attack introduced in \autoref{sec:meth:imperc}. The adversarial examples found by $L_\infty$ attack, though having small $L_\infty$ norm, have smaller structural similarity (SSIM) and larger color difference (CIEDE2000) compared to imperceptible attacks. Non-adaptive imperceptible attack result in smaller $L_0$, $L_2$, SSIM and CIEDE2000 measures, indicating smaller perceptual distance between original and adversarial images, however, it often significantly reduces the attack success rate. The adaptive imperceptible attack finds a balance between the attack success rate and the imperceptibility of the perturbation, which generates adversarial examples with slightly larger perceptual distance to the original image comparing with non-adaptive imperceptible attack, but greatly improves the attack success rate.

\begin{table}[!t]
\centering
\resizebox{\columnwidth}{!}{
\begin{tabular}{|l||l|l|l|l|l|l|l|}
\hline
model & Attack & Attack Success Rate & $L_0$ & $L_2$ & $L_\infty$ & SSIM & CIEDE2000 \\ \hline\hline
\multirow{3}{*}{M1} & $L_\infty$ & 100\% & 592.04 & 2.21 & \textbf{0.06} & 0.82 & - \\ \cline{2-8} 
 & Imperc & 24\% & 140.71 & \textbf{1.44} & 0.13 & \textbf{0.98} & - \\ \cline{2-8} 
 & Imperc-Adap & 60\% & \textbf{133.63} & 4.36 & 0.34 & 0.93 & - \\ \hline\hline
\multirow{3}{*}{M2} & $L_\infty$ & 100\% & 517.82 & 9.47 & 0.15 & 0.71 & - \\ \cline{2-8} 
 & Imperc & 14\% & 145.79 & \textbf{1.73} & \textbf{0.13} & \textbf{0.97} & - \\ \cline{2-8} 
 & Imperc-Adap & 52\% & \textbf{136.56} & 5.56 & 0.41 & 0.91 & - \\ \hline\hline
\multirow{3}{*}{C3} & $L_\infty$ & 100\% & 2815.62 & 0.81 & \textbf{0.01} & 0.98 & 70.50 \\ \cline{2-8} 
 & Imperc & 89\% & 2759.44 & \textbf{0.62} & 0.02 & \textbf{0.99} & \textbf{58.60} \\ \cline{2-8} 
 & Imperc-Adap & 99\% & \textbf{2731.90} & 0.94 & 0.03 & \textbf{0.99} & 68.71 \\ \hline\hline
\multirow{3}{*}{C4} & $L_\infty$ & 100\% & 3051.14 & 4.66 & \textbf{0.03} & 0.94 & 152.87 \\ \cline{2-8} 
 & Imperc & 58\% & 2937.72 & \textbf{1.54} & \textbf{0.04} & \textbf{0.98} & \textbf{92.23} \\ \cline{2-8} 
 & Imperc-Adap & 93\% & \textbf{2900.11} & 4.02 & 0.11 & 0.96 & 137.07 \\ \hline\hline
\multirow{3}{*}{S1} & $L_\infty$ & 100\% & 27236.80 & 0.74 & \textbf{0.01} & 0.99 & 151.30 \\ \cline{2-8} 
 & Imperc & 100\% & \textbf{24803.66} & \textbf{0.72} & \textbf{0.01} & \textbf{1.00} & \textbf{97.43} \\ \cline{2-8} 
 & Imperc-Adap & 100\% & \textbf{24803.66} & \textbf{0.72} & \textbf{0.01} & \textbf{1.00} & 97.50 \\ \hline
\end{tabular}
}
\caption{Different Measures on the adversarial examples generated by $L_\infty$ attack and Imperceptible attack.} 
\label{Tab:imperc}
\end{table}

\subsection{Discussion on imperceptibility}

The difficulty in attaining imperceptible attack depends on the robustness of the model and the input region allowed for perturbation. When the model has weak robustness and the region allowed for perturbation is large, it can be easy to derive an imperceptible attack, even without considering variance map. So, in \autoref{Tab:imperc}, we can find that in some cases, e.g. C3 and S1, the SSIM measures on the $L_\infty$ attack is also high, which means the adversarial perturbations generated by $L_\infty$ attack can also be quite imperceptible. The advantage of our variance map based imperceptible attack is not well demonstrated in such cases. Our method is more powerful when the model has relatively strong robustness, e.g., M1 and M2 in \autoref{Tab:imperc}, or when the perturbation is limited to a small region where normal $L_p$-norm based attacks can not achieve imperceptibility.

\section{Evaluation of Importance Map based Vulnerable Region Estimation}
\label{sec:eval:vulnerability}

We evaluate the effectiveness of our proposed importance map based method for finding the vulnerable region to adversarial attack. We estimate for an input image $x \in \mathbb{R}^{d}$ the vulnerability of a region specified by a binary mask $\omega \in {\{0, 1\}}^{d}$ using the minimal $L_\infty$ norm of the adversarial perturbation on this region, i.e., the robust radius of the region. The robust radius is estimated using our $L_\infty$ attack.

We conducted experiment by focusing on imposing rectangular regions of fixed 10 $\times$ 10 size on different datasets and models. To determine the robust radius within such a rectangular region for an image, We employed sliding window technique, sliding the rectangular region across the entire image and applying our $L_{\infty}$ attack on the region, and estimating the minimum robust radius among all the adversarial perturbations found. The final minimum robust radius is denoted as $r_{min}$. 

Next we computed the robust radius $r_h$ from the regions {\em selected by the importance map} generated using different heuristics $h$, and computed the ratio $ratio_h=\frac{r_h}{r_{min}}$ for each heuristic; this ratio illustrates the effectiveness of a heuristic in identifying regions with high vulnerability. We considered four different heuristics with respect to four importance maps, the GradCAM map, the GradCAM++ map, the IntergratedGradients+SmoothGrad map (IG+S) and the IntergratedGradients+SmoothGrad map with correction coefficient (IG+S($\beta$)). 

GradCAM \cite{selvaraju2016grad} is a method that uses gradient coming back to last Convolutional layer of CNN to assign importance weights to input pixels, with the same objective of Integrated Gradients, it also can be used to generate the importance map. GradCAM++ \cite{chattopadhay2018grad} is an improved version of GradCAM with similar usage. 
We compare the average $ratio_h$ over 100 image samples on each dataset and model (10 image samples for STL10 due to time efficiency) for the different heuristics, as shown in \autoref{Tab:region}. When no adversarial example is found in a region, the robust radius is recorded as 1.0, which is the length of the valid range of pixel value.

\begin{table}[t]
\centering
\resizebox{\columnwidth}{!}{
\begin{tabular}{|c|c|c|c|c|c|}
\hline
& $r_{min}$ & $ratio_{GradCAM}$ & $ratio_{GradCAM++}$ & $ratio_{IG+S}$ (ours) & $ratio_{IG+S(\beta)}$ (ours) \\ \hline
M1 & 0.2060 & - & - & \textbf{1.28} & 1.30 \\ \hline
M2 & 0.2610 & 2.16 & 1.97 & 1.21 & \textbf{1.19} \\ \hline
C1 & 0.05169 & - & - & 1.24 & \textbf{1.21} \\ \hline
C2 & 0.05595 & 2.95 & 3.76 & 1.80 & \textbf{1.65} \\ \hline
C3 & 0.05304 & 2.70 & 2.42 & 2.38 & \textbf{2.31} \\ \hline
C4 & 0.1046 & 3.16 & 4.61 & 1.52 & \textbf{1.46} \\ \hline
S1 & 0.1105 & 11.96 & 12.80 & \textbf{2.40} & 3.04 \\ \hline
\end{tabular}
}
\caption[Average $r_{min}$ and average $ratio_h$ for different heuristics.]{\textbf{Average $r_{min}$ and average $ratio_h$ for different heuristics.} GradCAM and GradCAM++ only works for Convolutional Networks, so not applicable to M1 and C1.}
\label{Tab:region}
\end{table}

We find that our method often select relatively good regions in affordable time. For C4 (image size 32 * 32), exhaustive applying $L_\infty$ attack on a sliding window on one image costs more than 5,000 seconds, while our importance map method costs 34 seconds. For S1 (image size 96 * 96), exhaustive search on one image costs more than 50,000 seconds, while our importance map method costs 28 seconds. The result also shows that importance map generated using Integrated gradient with SmoothGrad is good in estimating vulnerability, which is demonstrated by the comparison with GradCAM and GradCAM++.

We also experimented on improving our method by calculating on more regions, instead of selecting only the top 1 region with respect to importance score, we try to find the top k candidates and then selects the one actually returns smaller distance. We experiment on the IG+S($\beta$) map we proposed, and show how the $ratio_{IG+S(\beta)}$ and time cost change with the number of candidates k, as shown in \autoref{eval:fig:topk} and \autoref{eval:fig:topk_time}.

\begin{figure}[t]
  \centering
  \begin{minipage}[t]{.45\linewidth}
    \centering
    \includegraphics[width=\linewidth]{\Pic{png}{topk}}
    \caption[Change of $ratio_{IG+S(\beta)}$ with respect to k.]{Change of $ratio_{IG+S(\beta)}$ with respect to number of candidates k.}
    \label{eval:fig:topk}
  \end{minipage}
  \hspace*{2em}
  \begin{minipage}[t]{.45\linewidth}
    \centering
    \includegraphics[width=\linewidth]{\Pic{png}{topk_time}}
    \caption[Change of time cost with respect to k.]{Change of time cost with respect to number of candidates k.}
    \label{eval:fig:topk_time}
  \end{minipage}
\end{figure}

The $ratio_{IG+S(\beta)}$ can be largely reduced as the growing of k, while the time cost grows linearly, which suggests that we can select a suitable k (e.g. k=20) to balance the computing efficiency and the estimation accuracy of vulnerability.

\subsection{Discussion on the generality of regions}

Although we used a set of rectangular regions in our experiments to show the effective of our method, we can actually use regions of arbitrary shape: it can be any subset of all pixels in the image, and the set of pixels may or may not be clustered together. For example, it is possible to consider the set of regions each containing exactly $M$ pixels, where $M = 1, 2, ..., N$, where $N$ is the total number of pixels in the image. Such a set has cardinality  $N \choose M$, which can be quite large. Nevertheless, using our importance map based method, the most vulnerable region can be determined by simply selecting the $M$ pixels with the highest importance score.

\section{Evaluation of System Usability}

To evaluate the usability of our system. We created some instructions for testing our system, detailed in \autoref{append:instructions}, and asked 4 volunteers to use our system and fill out the Post-Study System Usability Questionnaire (PSSUQ) \cite{lewis1992psychometric}, which is widely used to measure users’ perceived satisfaction of a website, software, system or product at the end of a study. We used the third version of PSSUQ, which consists of 16 questions and follows a 7-point Likert Scale (+ NA option) between Strongly Agree to Strongly Disagree, as shown in \autoref{eval:fig:PSSUQ}.
\begin{figure}[h]
  \centering
  \includegraphics[width=\linewidth]{\Pic{png}{PSSUQ}}
  \vspace{\BeforeCaptionVSpace}
  \caption{The Post-Study System Usability Questionnaire (Version 3) \cite{lewis1992psychometric}}
  \label{eval:fig:PSSUQ}
\end{figure}

The overall result is calculated by averaging the scores from the 7 points of the scale. PSSUQ also has 3 sub-scales, namely system usefulness, information quality, and interface quality.
\begin{itemize}
    \item Overall: the average scores of questions 1 to 16
    \item System Usefulness (SYSUSE): the average scores of questions 1 to 6
    \item Information Quality (INFOQUAL): the average scores of questions 7 to 12
    \item Interface Quality (INTERQUAL): the average scores of questions 13 to 15
\end{itemize}

PSSUQ score starts with 1 (strongly agree) and ends with 7 (strongly disagree), with 4 being neutral. The lower the score, the better the performance and satisfaction. However, a score below 4 does not indicate that the system have performed above average. To help interpreting the PSSUQ scores, Sauro and Lewis \cite{sauro2016quantifying} analyzed the means of PSSUQ scores with data collected from 21 studies and 210 participants, as shown below:
\begin{itemize}
    \item SYSUSE: 2.80
    \item INFOQUAL: 3.02
    \item INTERQUAL: 2.49
    \item Overall: 2.82
\end{itemize}

The PSSUQ scores of our system calculated base on the questionnaire results we collected are:
\begin{itemize}
    \item SYSUSE: 1.71
    \item INFOQUAL: 1.92
    \item INTERQUAL: 1.92
    \item Overall: 1.81
\end{itemize}
which shows that the usability of our system is relatively good.

\section{Summary}

In this chapter, we evaluated the effectiveness of our system in three tasks: $L_\infty$ attack, imperceptible attack, and vulnerable region estimation. The experiment results showed that our method finds close or smaller $L_\infty$-norm adversarial examples compared with some state of the art $L_\infty$ attacks. We showed our system is able to generate more imperceptible adversarial perturbations compared with normal $L_\infty$ attack, and the adaptive loosening of the constraint can successfully increase the Attack Success Rate. We also showed that our proposed importance map method based on IntergratedGradients and SmoothGrad is able to identified vulnerable regions in the image, and our proposed correction coefficient can be used to improve the accuracy of vulnerability estimation. Finally, we evaluated the usability of our system based on users' feedback in the Post-Study System Usability Questionnaire, and found that our system performs above the average.

\chapter{Conclusion and Future Work}
\label{ch:concl}

\section{Conclusion}

We propose \attackname{}, a flexible adversarial perturbation generation system for deep learning models. We utilize various techniques in our system to generate adversarial perturbations with minimal $L_\infty$ norm, and provide flexibility with respect to regional and imperceptible adversarial perturbations. Different from many previously proposed $L_\infty$ attacks which perturb the whole inputs indiscriminately, we propose to use mask-constraints to generalize our attack to more scenarios, e.g. ``physical-world''. We show that out system is suitable for modeling various kinds of attacks, like imperceptible attack and regional attack. We also proposed variance map and importance map based methods to automatically generate imperceptible perturbation and approximately estimate the vulnerability of pixels/regions in an image. Extensive experiments show that our system is comparable to state of the art methods in terms of $L_\infty$ attack, is effective on generating imperceptible and regional adversarial perturbations, and can identify regions with high vulnerability. User studies of our system showed that our system also has good usability. 

\section{Future Research Directions}

Research on understanding robustness and adversarial examples of neural networks is still in its infancy. In the following, we discuss a few future research directions.

\subsection{Attack For Image Segmentation/Object Detection}

Besides image classification, image segmentation and object detection are among the mainstream problems in modern deep learning applications \cite{long2015fully,he2017mask,girshick2014rich,girshick2015fast,ren2015faster}. While many efforts on adversarial attack and defence have been spent on image classification networks, there are not enough attention drawn to image segmentation and object detection networks. Nevertheless, image segmentation and object detection are closely related to some safe-critical systems, e.g. face detection \cite{hjelmaas2001face}, medical imaging \cite{suzuki2017overview} and autonomous driving \cite{grigorescu2020survey}. Research on attack for image segmentation/object detection are certainly needed to identify potential threats. It would be interesting to see if regional and imperceptible attack can be achieved in these scenarios.

\subsection{Interpretable Attack}

While our method of identifying vulnerable region can provide information about the weakness of a model, specifically, which pixels are unstable when facing adversarial attack, the perturbations generated by the attack methods remain uninterpretable to human, which means the cause and effect can not be determined. Why such adversarial perturbations that are incomprehensible to humans can cause the model to produce erroneous output? Is it possible to generate human comprehensible perturbations? It would be interesting to understand how the findings of Ilyas et al. \cite{ilyas2019adversarial} are related to such perturbations. By answering these questions, we might be able to better understand the threats faced by our machine learning models, and thus push the related research towards a safer and more trustworthy way.

\vspace{2em}
We believe the above (but not limited to) future research directions will advance the technology presented in this thesis and contribute to academia and industry.

\bookmarksetup{startatroot}
\printbibliography[heading=bibintoc]

@article{voulodimos2018deep,
  title={Deep learning for computer vision: A brief review},
  author={Voulodimos, Athanasios and Doulamis, Nikolaos and Doulamis, Anastasios and Protopapadakis, Eftychios},
  journal={Computational intelligence and neuroscience},
  volume={2018},
  year={2018},
  publisher={Hindawi}
}

@article{otter2020survey,
  title={A survey of the usages of deep learning for natural language processing},
  author={Otter, Daniel W and Medina, Julian R and Kalita, Jugal K},
  journal={IEEE transactions on neural networks and learning systems},
  volume={32},
  number={2},
  pages={604--624},
  year={2020},
  publisher={IEEE}
}

@article{purwins2019deep,
  title={Deep learning for audio signal processing},
  author={Purwins, Hendrik and Li, Bo and Virtanen, Tuomas and Schl{\"u}ter, Jan and Chang, Shuo-Yiin and Sainath, Tara},
  journal={IEEE Journal of Selected Topics in Signal Processing},
  volume={13},
  number={2},
  pages={206--219},
  year={2019},
  publisher={IEEE}
}

@article{szegedy2013intriguing,
  title={Intriguing properties of neural networks},
  author={Szegedy, Christian and Zaremba, Wojciech and Sutskever, Ilya and Bruna, Joan and Erhan, Dumitru and Goodfellow, Ian and Fergus, Rob},
  journal={arXiv preprint arXiv:1312.6199},
  year={2013}
}

@article{goodfellow2014explaining,
  title={Explaining and harnessing adversarial examples},
  author={Goodfellow, Ian J and Shlens, Jonathon and Szegedy, Christian},
  journal={arXiv preprint arXiv:1412.6572},
  year={2014}
}

@article{madry2017towards,
  title={Towards deep learning models resistant to adversarial attacks},
  author={Madry, Aleksander and Makelov, Aleksandar and Schmidt, Ludwig and Tsipras, Dimitris and Vladu, Adrian},
  journal={arXiv preprint arXiv:1706.06083},
  year={2017}
}

@inproceedings{moosavi2016deepfool,
  title={Deepfool: a simple and accurate method to fool deep neural networks},
  author={Moosavi-Dezfooli, Seyed-Mohsen and Fawzi, Alhussein and Frossard, Pascal},
  booktitle={Proceedings of the IEEE conference on computer vision and pattern recognition},
  pages={2574--2582},
  year={2016}
}

@inproceedings{carlini2017towards,
  title={Towards evaluating the robustness of neural networks},
  author={Carlini, Nicholas and Wagner, David},
  booktitle={2017 ieee symposium on security and privacy (sp)},
  pages={39--57},
  year={2017},
  organization={IEEE}
}

@inproceedings{papernot2016limitations,
  title={The limitations of deep learning in adversarial settings},
  author={Papernot, Nicolas and McDaniel, Patrick and Jha, Somesh and Fredrikson, Matt and Celik, Z Berkay and Swami, Ananthram},
  booktitle={2016 IEEE European symposium on security and privacy (EuroS\&P)},
  pages={372--387},
  year={2016},
  organization={IEEE}
}

@article{krizhevsky2012imagenet,
  title={Imagenet classification with deep convolutional neural networks},
  author={Krizhevsky, Alex and Sutskever, Ilya and Hinton, Geoffrey E},
  journal={Advances in neural information processing systems},
  volume={25},
  year={2012}
}

@inproceedings{he2016deep,
  title={Deep residual learning for image recognition},
  author={He, Kaiming and Zhang, Xiangyu and Ren, Shaoqing and Sun, Jian},
  booktitle={Proceedings of the IEEE conference on computer vision and pattern recognition},
  pages={770--778},
  year={2016}
}

@inproceedings{huang2017densely,
  title={Densely connected convolutional networks},
  author={Huang, Gao and Liu, Zhuang and Van Der Maaten, Laurens and Weinberger, Kilian Q},
  booktitle={Proceedings of the IEEE conference on computer vision and pattern recognition},
  pages={4700--4708},
  year={2017}
}

@inproceedings{long2015fully,
  title={Fully convolutional networks for semantic segmentation},
  author={Long, Jonathan and Shelhamer, Evan and Darrell, Trevor},
  booktitle={Proceedings of the IEEE conference on computer vision and pattern recognition},
  pages={3431--3440},
  year={2015}
}

@inproceedings{he2017mask,
  title={Mask r-cnn},
  author={He, Kaiming and Gkioxari, Georgia and Doll{\'a}r, Piotr and Girshick, Ross},
  booktitle={Proceedings of the IEEE international conference on computer vision},
  pages={2961--2969},
  year={2017}
}

@inproceedings{girshick2014rich,
  title={Rich feature hierarchies for accurate object detection and semantic segmentation},
  author={Girshick, Ross and Donahue, Jeff and Darrell, Trevor and Malik, Jitendra},
  booktitle={Proceedings of the IEEE conference on computer vision and pattern recognition},
  pages={580--587},
  year={2014}
}

@inproceedings{girshick2015fast,
  title={Fast r-cnn},
  author={Girshick, Ross},
  booktitle={Proceedings of the IEEE international conference on computer vision},
  pages={1440--1448},
  year={2015}
}

@article{ren2015faster,
  title={Faster r-cnn: Towards real-time object detection with region proposal networks},
  author={Ren, Shaoqing and He, Kaiming and Girshick, Ross and Sun, Jian},
  journal={Advances in neural information processing systems},
  volume={28},
  year={2015}
}

@inproceedings{katz2017reluplex,
  title={Reluplex: An efficient SMT solver for verifying deep neural networks},
  author={Katz, Guy and Barrett, Clark and Dill, David L and Julian, Kyle and Kochenderfer, Mykel J},
  booktitle={International conference on computer aided verification},
  pages={97--117},
  year={2017},
  organization={Springer}
}

@article{zhai2020macer,
  title={Macer: Attack-free and scalable robust training via maximizing certified radius},
  author={Zhai, Runtian and Dan, Chen and He, Di and Zhang, Huan and Gong, Boqing and Ravikumar, Pradeep and Hsieh, Cho-Jui and Wang, Liwei},
  journal={arXiv preprint arXiv:2001.02378},
  year={2020}
}

@inproceedings{sharif2018suitability,
  title={On the suitability of lp-norms for creating and preventing adversarial examples},
  author={Sharif, Mahmood and Bauer, Lujo and Reiter, Michael K},
  booktitle={Proceedings of the IEEE Conference on Computer Vision and Pattern Recognition Workshops},
  pages={1605--1613},
  year={2018}
}

@article{liu2010just,
  title={Just noticeable difference for images with decomposition model for separating edge and textured regions},
  author={Liu, Anmin and Lin, Weisi and Paul, Manoranjan and Deng, Chenwei and Zhang, Fan},
  journal={IEEE Transactions on Circuits and Systems for Video Technology},
  volume={20},
  number={11},
  pages={1648--1652},
  year={2010},
  publisher={IEEE}
}

@article{lecun2015deep,
  title={Deep learning},
  author={LeCun, Yann and Bengio, Yoshua and Hinton, Geoffrey},
  journal={nature},
  volume={521},
  number={7553},
  pages={436--444},
  year={2015},
  publisher={Nature Publishing Group}
}

@article{bre2018prediction,
  title={Prediction of wind pressure coefficients on building surfaces using artificial neural networks},
  author={Bre, Facundo and Gimenez, Juan M and Fachinotti, V{\'\i}ctor D},
  journal={Energy and Buildings},
  volume={158},
  pages={1429--1441},
  year={2018},
  publisher={Elsevier}
}

@article{lu2017no,
  title={No need to worry about adversarial examples in object detection in autonomous vehicles},
  author={Lu, Jiajun and Sibai, Hussein and Fabry, Evan and Forsyth, David},
  journal={arXiv preprint arXiv:1707.03501},
  year={2017}
}

@article{yosinski2015understanding,
  title={Understanding neural networks through deep visualization},
  author={Yosinski, Jason and Clune, Jeff and Nguyen, Anh and Fuchs, Thomas and Lipson, Hod},
  journal={arXiv preprint arXiv:1506.06579},
  year={2015}
}

@article{wang2004image,
  title={Image quality assessment: from error visibility to structural similarity},
  author={Wang, Zhou and Bovik, Alan C and Sheikh, Hamid R and Simoncelli, Eero P},
  journal={IEEE transactions on image processing},
  volume={13},
  number={4},
  pages={600--612},
  year={2004},
  publisher={IEEE}
}

@article{luo2001development,
  title={The development of the CIE 2000 colour-difference formula: CIEDE2000},
  author={Luo, M Ronnier and Cui, Guihua and Rigg, Bryan},
  journal={Color Research \& Application: Endorsed by Inter-Society Color Council, The Colour Group (Great Britain), Canadian Society for Color, Color Science Association of Japan, Dutch Society for the Study of Color, The Swedish Colour Centre Foundation, Colour Society of Australia, Centre Fran{\c{c}}ais de la Couleur},
  volume={26},
  number={5},
  pages={340--350},
  year={2001},
  publisher={Wiley Online Library}
}

@inproceedings{chattopadhay2018grad,
  title={Grad-cam++: Generalized gradient-based visual explanations for deep convolutional networks},
  author={Chattopadhay, Aditya and Sarkar, Anirban and Howlader, Prantik and Balasubramanian, Vineeth N},
  booktitle={2018 IEEE winter conference on applications of computer vision (WACV)},
  pages={839--847},
  year={2018},
  organization={IEEE}
}

@inproceedings{sharif2016accessorize,
  title={Accessorize to a crime: Real and stealthy attacks on state-of-the-art face recognition},
  author={Sharif, Mahmood and Bhagavatula, Sruti and Bauer, Lujo and Reiter, Michael K},
  booktitle={Proceedings of the 2016 acm sigsac conference on computer and communications security},
  pages={1528--1540},
  year={2016}
}

@article{singh2019abstract,
  title={An abstract domain for certifying neural networks},
  author={Singh, Gagandeep and Gehr, Timon and P{\"u}schel, Markus and Vechev, Martin},
  journal={Proceedings of the ACM on Programming Languages},
  volume={3},
  number={POPL},
  pages={1--30},
  year={2019},
  publisher={ACM New York, NY, USA}
}

@article{brown2017adversarial,
  title={Adversarial patch},
  author={Brown, Tom B and Man{\'e}, Dandelion and Roy, Aurko and Abadi, Mart{\'\i}n and Gilmer, Justin},
  journal={arXiv preprint arXiv:1712.09665},
  year={2017}
}

@article{brendel2019accurate,
  title={Accurate, reliable and fast robustness evaluation},
  author={Brendel, Wieland and Rauber, Jonas and K{\"u}mmerer, Matthias and Ustyuzhaninov, Ivan and Bethge, Matthias},
  journal={Advances in neural information processing systems},
  volume={32},
  year={2019}
}

@misc{brendel_2020, 
  title={Accurate, reliable and fast robustness evaluation}, 
  url={https://medium.com/bethgelab/accurate-reliable-and-fast-robustness-evaluation-4e2a5ab43521},
  journal={Medium}, 
  author={Brendel, Wieland}, 
  year={2020}, 
  month={Jan}
}

@inproceedings{luo2018towards,
  title={Towards imperceptible and robust adversarial example attacks against neural networks},
  author={Luo, Bo and Liu, Yannan and Wei, Lingxiao and Xu, Qiang},
  booktitle={Proceedings of the AAAI Conference on Artificial Intelligence},
  volume={32},
  number={1},
  year={2018}
}

@inproceedings{croce2019sparse,
  title={Sparse and imperceivable adversarial attacks},
  author={Croce, Francesco and Hein, Matthias},
  booktitle={Proceedings of the IEEE/CVF International Conference on Computer Vision},
  pages={4724--4732},
  year={2019}
}

@inproceedings{karmon2018lavan,
  title={Lavan: Localized and visible adversarial noise},
  author={Karmon, Danny and Zoran, Daniel and Goldberg, Yoav},
  booktitle={International Conference on Machine Learning},
  pages={2507--2515},
  year={2018},
  organization={PMLR}
}

@article{dia2021localized,
  title={Localized Uncertainty Attacks},
  author={Dia, Ousmane Amadou and Karaletsos, Theofanis and Hazirbas, Caner and Ferrer, Cristian Canton and Kabul, Ilknur Kaynar and Meijer, Erik},
  journal={arXiv preprint arXiv:2106.09222},
  year={2021}
}

@article{bai2021inconspicuous,
  title={Inconspicuous Adversarial Patches for Fooling Image Recognition Systems on Mobile Devices},
  author={Bai, Tao and Luo, Jinqi and Zhao, Jun},
  journal={IEEE Internet of Things Journal},
  year={2021},
  publisher={IEEE}
}

@article{selvaraju2016grad,
  title={Grad-CAM: Why did you say that?},
  author={Selvaraju, Ramprasaath R and Das, Abhishek and Vedantam, Ramakrishna and Cogswell, Michael and Parikh, Devi and Batra, Dhruv},
  journal={arXiv preprint arXiv:1611.07450},
  year={2016}
}

@inproceedings{sundararajan2017axiomatic,
  title={Axiomatic attribution for deep networks},
  author={Sundararajan, Mukund and Taly, Ankur and Yan, Qiqi},
  booktitle={International conference on machine learning},
  pages={3319--3328},
  year={2017},
  organization={PMLR}
}

@article{smilkov2017smoothgrad,
  title={Smoothgrad: removing noise by adding noise},
  author={Smilkov, Daniel and Thorat, Nikhil and Kim, Been and Vi{\'e}gas, Fernanda and Wattenberg, Martin},
  journal={arXiv preprint arXiv:1706.03825},
  year={2017}
}

@article{lecun-mnisthandwrittendigit-2010,
  author = {LeCun, Yann and Cortes, Corinna},
  howpublished = {http://yann.lecun.com/exdb/mnist/},
  lastchecked = {2016-01-14 14:24:11},
  title = {{MNIST} handwritten digit database},
  url = {http://yann.lecun.com/exdb/mnist/},
  username = {mhwombat},
  year = 2010
}

@article{krizhevsky2009learning,
  title={Learning multiple layers of features from tiny images},
  author={Krizhevsky, Alex and Hinton, Geoffrey and others},
  year={2009},
  publisher={Citeseer}
}

@inproceedings{coates2011analysis,
  title={An analysis of single-layer networks in unsupervised feature learning},
  author={Coates, Adam and Ng, Andrew and Lee, Honglak},
  booktitle={Proceedings of the fourteenth international conference on artificial intelligence and statistics},
  pages={215--223},
  year={2011},
  organization={JMLR Workshop and Conference Proceedings}
}

@inproceedings{croce2020minimally,
  title={Minimally distorted adversarial examples with a fast adaptive boundary attack},
  author={Croce, Francesco and Hein, Matthias},
  booktitle={International Conference on Machine Learning},
  pages={2196--2205},
  year={2020},
  organization={PMLR}
}

@article{pintor2021fast,
  title={Fast minimum-norm adversarial attacks through adaptive norm constraints},
  author={Pintor, Maura and Roli, Fabio and Brendel, Wieland and Biggio, Battista},
  journal={Advances in Neural Information Processing Systems},
  volume={34},
  year={2021}
}

@inproceedings{mirman2018differentiable,
  title={Differentiable abstract interpretation for provably robust neural networks},
  author={Mirman, Matthew and Gehr, Timon and Vechev, Martin},
  booktitle={International Conference on Machine Learning},
  pages={3578--3586},
  year={2018},
  organization={PMLR}
}

@article{hjelmaas2001face,
  title={Face detection: A survey},
  author={Hjelm{\aa}s, Erik and Low, Boon Kee},
  journal={Computer vision and image understanding},
  volume={83},
  number={3},
  pages={236--274},
  year={2001},
  publisher={Elsevier}
}

@article{suzuki2017overview,
  title={Overview of deep learning in medical imaging},
  author={Suzuki, Kenji},
  journal={Radiological physics and technology},
  volume={10},
  number={3},
  pages={257--273},
  year={2017},
  publisher={Springer}
}

@article{grigorescu2020survey,
  title={A survey of deep learning techniques for autonomous driving},
  author={Grigorescu, Sorin and Trasnea, Bogdan and Cocias, Tiberiu and Macesanu, Gigel},
  journal={Journal of Field Robotics},
  volume={37},
  number={3},
  pages={362--386},
  year={2020},
  publisher={Wiley Online Library}
}

@article{ilyas2019adversarial,
  title={Adversarial examples are not bugs, they are features},
  author={Ilyas, Andrew and Santurkar, Shibani and Tsipras, Dimitris and Engstrom, Logan and Tran, Brandon and Madry, Aleksander},
  journal={Advances in neural information processing systems},
  volume={32},
  year={2019}
}

@book{pugh2002real,
  title={Real mathematical analysis},
  author={Pugh, Charles Chapman and Pugh, CC},
  volume={2011},
  year={2002},
  publisher={Springer}
}

@inproceedings{lewis1992psychometric,
  title={Psychometric evaluation of the post-study system usability questionnaire: The PSSUQ},
  author={Lewis, James R},
  booktitle={Proceedings of the human factors society annual meeting},
  volume={36},
  number={16},
  pages={1259--1260},
  year={1992},
  organization={Sage Publications Sage CA: Los Angeles, CA}
}

@book{sauro2016quantifying,
  title={Quantifying the user experience: Practical statistics for user research},
  author={Sauro, Jeff and Lewis, James R},
  year={2016},
  publisher={Morgan Kaufmann}
}

\appendix
\chapter{Distance Metric}
\label{append:metric}

\section{SSIM}
The structural similarity index measure (SSIM) is a perception-based metric to measure the similarity between two images. Given two images $x$ and $y$, the structural similarity of the two images can be calculated as follows:
\begin{equation}
  SSIM(x,y) = \frac{(2\mu_x\mu_y + C_1) + (2 \sigma _{xy} + C_2)} {(\mu_x^2 + \mu_y^2+C_1) (\sigma_x^2 + \sigma_y^2+C_2)}
  \label{eq:SSMI}
\end{equation}
where $\mu_x$ is the mean of $x$, $\mu_y$ is the mean of $y$, $\sigma_x^2$ is the variance of $x$, $\sigma_y^2$ is the variance of $y$, $\sigma _{xy}$ is the covariance of $x$ and $y$, $C_1=(k_1L)^2$, $C_2=(k_2L)^2$ are constants used to maintain stability, where $L$ is the dynamic range of the pixel-values (typically $2^{\#bits\ per\ pixel}-1$), $k_1=0.01$ and $k_2=0.03$ by default. Structural similarity ranges from -1 to 1. When two images are identical, the value of SSIM is equal to 1.

\section{CIEDE2000}
CIEDE2000 is the latest $\Delta E^*$ color difference formula developed by the International Commission on Illumination (CIE). The pixel-wise perceptual color distance is calculated as:
\begin{equation}
  \Delta E^*_{00} = \sqrt{(\frac{\Delta L'}{k_LS_L})^2+(\frac{\Delta C'}{k_CS_C})^2+(\frac{\Delta H'}{k_HS_H})^2+R_T\frac{\Delta C'}{k_CS_C}\frac{\Delta H'}{k_HS_H}}
  \label{eq:CIEDE2000}
\end{equation}
where $\Delta L'$, $\Delta C'$, $\Delta H'$ denotes the distance between pixel values in the three channels, L (lightness), C (chroma) and H (hue), $S_L$, $S_C$, $S_H$ and $R_T$ are weighting functions used to compensate color space uniformity, $k_L$, $k_C$, $k_H$ are weighting factors (usually unity) relative to experimental conditions.
\chapter{System test Instructions}
\label{append:instructions}

Hi, thank you for willing to test our adversarial attack system!

Given an image classifier and an input image, an adversarial attack generates small adversarial perturbation on the input image to make the classifier produce wrong output.

The whole process should take you about 15 minutes. After you finish it, we would like you to provide feedback according to you experience during the testing.

To begin testing follow these steps:

\begin{enumerate}

    \item Setup the environment on your local machine.
    \begin{enumerate}
        \item Ensure that you have a python>=3.7 installed as the default python version on your local machine.
        \item Go to a local directory where you want to install our system.
        \item Clone the github repository by running
        
        \textbf{git clone https://github.com/SUSTC11612405/ExploreADV.git}
        \item cd into the source root
        
        \textbf{cd ExploreADV}
        \item Setup a virtual environment (Optional).
        
        \textbf{pip3 install virtualenv}
        
        \textbf{python3 -m virtualenv venv}
        
        \textbf{source venv/bin/activate}
        
        \item Install the dependencies.
        
        \textbf{pip3 install -r requirements.txt}
    \end{enumerate}
    
    \item Start running the system, you may need to download the datasets when you run the system for the first time.
    \begin{enumerate}
        \item Take a look a the usage
        
        \textbf{python run\_attack.py -h}
        \item Try to run the normal $L_\infty$ attack
        
        \textbf{python run\_attack.py}
        \item Try to run the attack on 30\% of the pixels
        
        \textbf{python run\_attack.py --ratio 0.3}
        \item Try to run the imperceptible attack
        
        \textbf{python run\_attack.py --imperceptible}
        \item Try to run attack on specified region using region selector. A GUI should show up for you to specify the region for attack, please feel free to click any button and play with the GUI.
        
        \textbf{python run\_attack.py --region select}
        \item Try to switch dataset and models, and explore any adversarial examples as you like. For example:
        
        \textbf{python run\_attack.py --dataset cifar10 --path\_model ./models/cifar10\_convBigRELU\_DiffAI.onnx --region select --imperceptible}
        
        \textbf{python run\_attack.py --region select --ratio 0.1}
    \end{enumerate}
    
    \item Congratulations on completing the tasks! Hope you had fun. Now, we would like to ask you to fill out the following PSSUQ questionnaire to evaluate our system. Thank you again for your effort!
    \begin{figure}[h]
      \centering
      \includegraphics[width=\linewidth]{\Pic{png}{PSSUQ}}
      \vspace{\BeforeCaptionVSpace}
      \caption{The Post-Study System Usability Questionnaire (Version 3)}
    \end{figure}
\end{enumerate}

\end{document}